\documentclass[preprint]{aastex}

\renewcommand{\sec}{\prime\prime}
\renewcommand{\min}{\prime}
\renewcommand{\deg}[0]{\circ}

\begin{document}

\title{Tidal Interaction of M32 and NGC~205 with M31: Surface
Photometry and Numerical Simulations\footnote{Observations carried out
at Kitt Peak National Observatory, National Optical Astronomy
Observatories, which is operated by the Association of Universities
for Research in Astronomy, Inc. under cooperative agreement with the
National Science Foundation.}}

\author{Philip I. Choi, Puragra Guhathakurta\\ UCO/Lick Observatory,
University of California, Santa Cruz, CA 95064\\ E-mail:
pchoi@ucolick.org, raja@ucolick.org}

\author{Kathryn V. Johnston\\ Van Vleck Observatory, Wesleyan
University, Middletown, CT 06459\\ E-mail: kvj@astro.wesleyan.edu}

\begin{abstract} 

We investigate the interaction history of the M31 sub-group by
comparing surface photometry of two of its satellites, M32 and
NGC~205, with N-body simulations of satellite destruction.
The recent discovery of a giant stream in the outer halo of M31,
apparently pointed in the direction of M32 and NGC~205, makes such
an investigation particularly relevant.
The observational component of this study is based on
$1.7^{\circ}\times5^{\circ}$ $B$- and $I$-band CCD mosaic images
centered on M31 and covering both satellites.  Standard
ellipse-fitting techniques are used to model and remove M31 disk light
and to perform surface photometry on the satellites to limiting
brightness levels of $[\mu_B,~\mu_I]=[27,~25]~{\rm mag~arcsec}^{-2}$,
corresponding to isophotal semi-major axis lengths of $r_{\rm
lim}^{\rm M32}=420^{\sec}$ (1.6~kpc) and $r_{\rm lim}^{\rm
NGC\,205}=720^{\sec}$ (2.7~kpc).  A hint of excess light in the outer
parts of M32 noted in earlier studies is confirmed; in particular,
clear evidence is seen for a sharp (upward) break in the surface
brightness profile at $r=150^{\sec}$ relative to a $r^{1/4}$ law that
fits the inner region of M32.  This break is accompanied by a steep
increase in isophotal ellipticity $\epsilon$ as well as position angle
$\phi^{\prime}$ twisting.  In addition to this excess, evidence is
seen for an inner downward break in the surface brightness profile at
$r=50^{\sec}$.  The robustness of the M32 isophotal features is
demonstrated through their: (1)~insensitivity to the details of
background subtraction; (2)~symmetry about M32's center; and
(3)~narrow range of $B-I$ color that is consistent with the interior
regions of M32 but not with M31 residual spiral arm/dust lane
features.  The study of NGC~205 reveals pronounced isophote twisting
at $r\sim300^{\sec}$ that is coincident with a subtle downward break
in the surface brightness profile, relative to an exponential law fit
to the inner region.

The simulation component of this project is based on the analysis of
single-component, spherical satellites that are being tidally
disrupted through interactions with their parent galaxy.  Generic
features of the simulations include an excess in the surface
brightness profile at large radii, a depletion zone at intermediate
radii, and isophotal elongation and twists that are coincident with
breaks in the brightness profile.  The two satellites, M32 and
NGC~205, display most of these features consistently across the $B$
and $I$ bands, strongly suggestive of tidal interaction and probable
stripping by M31.  We discuss what these observed features can tell us
about the satellites' orbital parameters and histories.  Specifically,
M32 is found to be on a highly eccentric orbit and away from
pericenter.  Investigating M32's unusual combination of high surface
brightness and low luminosity (the hallmark of compact ellipticals),
we make empirical estimates of the galaxy's intrinsic properties and
conclude that it is not likely to be the residual core of a
tidally-stripped normal elliptical galaxy as has been suggested, but
rather that its precursor was intrinsically compact.

\end{abstract}

\clearpage

\section{Introduction}
\label{intro}

Globular clusters and satellite galaxies serve as convenient tracers
of the mass distribution of their parent galaxy.  In theory, even a
few well-determined satellite orbits can be used to constrain the
gravitational potential field of the central galaxy \citep{evans+00}.
Unfortunately, direct measurement of orbital parameters --- the proper
motion, in particular --- is difficult.  It has long been believed
that observable signatures of tidal interaction in the satellites can
be used to determine at least some of these critical parameters.  For
instance, it was proposed that globular cluster profiles are limited
by the Galactic tidal field in which they are embedded
\citep{von_hoerner57} and that the anomalous properties of some
peculiar elliptical (E) galaxies could be the result of similar tidal
interactions \citep{king62,aw86}.  Since that time, there have been
numerous investigations into the dynamics of tidally-truncated
systems.  Objects such as globular clusters and compact elliptical (cE)
galaxies, of which M32 is a prototype, have been modeled with the King
modification of the von Hoerner tidal radius formula:

\begin{equation}
        r_{\rm tide,\,peri}= R_{\rm peri} \left[{ m_{\rm sat} \over
M_{\rm gal,\,peri} (e_{\rm orb}+3)}\right]^{1/3}
\label{rtide}
\end{equation}

\noindent where $r_{\rm tide,\,peri}$ is the tidal radius of the
satellite set at pericenter; $R_{\rm peri}$ is the satellite's
pericenter distance; $m_{\rm sat}$ and $M_{\rm gal,\,peri}$ are the
mass of the satellite galaxy and the mass of the parent galaxy
enclosed within the satellite's orbit, respectively; and $e_{\rm orb}$
is the satellite's orbital eccentricity.

Several studies have been based on the assumption that the limiting
radius of a truncated object corresponds to its tidal radius at
pericenter.  \citet{faber73} derived perigalacticon distances for a
sample of cE galaxies.  This was followed by more
ambitious attempts to constrain the orbital parameters of Galactic
globular clusters \citep{peterson74,innanen83} and M31 satellite
galaxies \citep{cepa_beckman88}.  These interpretations provided a
qualitative picture of the interactions; however, uncertainties in the
determination of the tidal radius prevented the accurate recovery of
orbital parameters.  In addition, the discovery of extra-tidal stars
around Galactic globular clusters \citep{g+95} and dwarf spheroidals
\citep{irwin_hatzidimitriou95,kuhn+96} complicated the notion of a
well-defined, observable tidal radius.  In response to these findings,
a slew of detailed numerical simulations emerged that modeled
extra-tidal features as well as extended tidal tails
\citep{oh+95,moore96,combes+99,johnston+99a}.  In turn, these
motivated further observational studies to more precisely characterize
both of these peripheral populations \citep{majewski+00,leon+00}.
Comparisons between observations and models are proving to be powerful
tools for probing the Galactic potential \citep{johnston+99b} and
determining satellite dark matter fractions, mass-loss rates (Johnston
et~al.\ 1999a), and orbital parameters.

The proximity of Galactic satellites makes detailed observations
possible but we are more or less limited to viewing them from within
the plane of their orbit.  External systems, while observationally
more challenging, offer the advantage of a global perspective on the
parent galaxy and a bird's-eye view of the satellites' orbits.  Our
nearest large galaxy neighbor, the Andromeda spiral (M31), has been
the subject of such studies for the last few decades
\citep{byrd79,sato_sawa86}.  Galaxy interactions in the M31 subgroup
have recently been in the limelight due to the discovery of a tidal
stream in the outer halo of M31 \citep{ibata+01} and hints of tidal
debris around its dwarf spheroidal satellites \citep{ostheimer+02}.
In this paper, we investigate signatures of tidal interaction in the
outskirts of the luminous M31 satellites, M32 and NGC~205.  This is
especially relevant since the Ibata et~al.\ stream lies, at least in
projection, along a line intersecting both M32 and NGC~205.  Our study
uses traditional integrated surface photometry techniques in contrast
to the star count analyses of Ibata et~al. and Ostheimer et~al..
Studies of satellite interactions well beyond the Local Group will,
into the foreseeable future, likely be restricted to the use of
surface photometry methods on relatively high surface brightness,
luminous satellites; thus, our work on M32 and NGC~205 may be viewed
as a pilot study for more distant systems.

In addition to probing the parent galaxy potential, it is interesting
to investigate the impact of tidal interactions on the morphology and
evolution of low-mass satellites.  The satellites M32 and NGC~205 represent
two distinct classes of low-mass galaxies, cEs and
dwarf ellipticals (dEs), respectively.  Normal E
galaxies are found to populate a region in luminosity ($L$),
surface brightness ($\mu$), and internal velocity dispersion
($\sigma$) space called the Fundamental Plane (FP).  In the $\mu$-$L$
projection of this space, the E galaxy population fainter than
$M_B\sim-18$ bifurcates into tracks of (1)~high surface brightness,
high-metallicity cEs and (2)~low surface brightness, low-metallicity
dEs \citep{kormendy85}.  There is no clear formation scenario unifying
these three classes of galaxies, and there has been a long-standing
debate about whether cEs or dEs represent the natural low-mass
extension of normal Es \citep{faber73,wg84,np87,bn90,kd89}.

As a class, cE galaxies have de~Vaucouleurs law $\mu$ profiles
like normal Es.  Furthermore, they occupy a region of structural parameter
space that is the direct low-luminosity
extrapolation of the E galaxy fundamental plane \citep{wg84,np87}, though
well separated from it.  On the other hand, the general proximity of cEs to
massive parent galaxies has led to speculation that cEs are formed
through the capture and tidal truncation of satellite galaxies
\citep{king_kiser73, tonry84, tonry87}.  The range of proposed cE
progenitors includes Es \citep{faber73}, S0s \citep{nieto90}, and
spirals \citep{bekki+01}.  Alternatively, \citet{burkert94a} has
proposed a model in which cEs are formed through a starburst and
subsequent violent collapse within the potential well of a massive
galaxy.  If cEs are the low-mass counterparts of normal Es,
their rarity \citep{zb98,dg98} and small range of absolute magnitudes
would imply a sharp turnover in the E galaxy mass function.

By contrast, dEs tend to: (1)~be fit by a King or exponential $\mu$
profile instead of a de~Vaucouleurs law and (2)~form a track
in $\mu$-$L$ space that is perpendicular to the classical E galaxy
track.  Such structural differences do not however rule out the possibility
of a connection between dE and E galaxies since different physical processes
may be at work in high versus low mass
galaxies.  The conventional wisdom regarding dE
formation has been that, given their low binding energies, they are
susceptible to supernova-driven galactic winds that regulate star
formation, expand the stellar component, and thereby produce diffuse
density profiles (Burkert 1994b and references therein).  Recent
simulations have contested these claims and suggested that mechanisms
such as galaxy harassment \citep{moore+96,mlk98,moore+99} and tidal
heating \citep{mayer+01a,mayer+01b} may be responsible for the
transformation of spiral and dwarf irregular galaxies into dEs.  Far
more numerous than cEs, dEs populate an wide range of absolute magnitudes
fainter than $M_B\sim-18.0$, beyond the sharp faint-end cutoff of the cE
luminosity function.  If dEs are the low-mass counterparts of normal Es, it
would imply that the low end of the E galaxy mass function has a smooth
extension.

Tidal interactions may also have some bearing on M32's unusual stellar
content.  Its stellar mix has been a controversial topic, with suggestions
ranging from a pure old population \citep{cole+98} to a single coeval
intermediate-age population \citep{va99,delBurgo+01}.  More plausibly, M32
seems to contain a small fraction of intermediate-age stars mixed in with an
underlying old population \citep{oconnell80,burstein+84,rose85,bas90,
davidge+00}.  Proposed theories for the origin of this secondary stellar
population invoke galaxy interactions as a trigger for star formation within
M32 or in the context of accretion of gaseous material.

The M31 satellites M32 and NGC~205 are thus good test subjects for
investigating the formation and evolution of these two classes of low-mass
early-type galaxies.  In this paper, a large-format CCD mosaic image is used
to carry out surface photometry of the satellites.  Earlier studies of these
systems have been plagued by large uncertainties in the measurement of their
faint outer isophotes: photographic studies \citep{deV53,hodge73} are
hampered by low-level plate-fog variations, while more recent CCD
observations \citep{kent87,peletier93} have limited fields of view making sky
subtraction problematic.  The situation is complicated by the fact that the
satellites' outer brightness profiles are contaminated by M31 disk light.
The large field of view of our CCD mosaic image makes global modeling and
subtraction of M31's disk light possible, thereby allowing for reliable
measurement of the satellites' faint isophotes.  These measurements are
compared to numerical simulations to place constraints on the orbital
parameters and mass-loss rates of the satellites and to estimate the
evolution of M32's luminosity and central surface brightness.
 
This paper is divided into the following sections.  A summary of the
observations and an overview of the basic data reduction procedure are given
in \S2.  The removal of ``background'' M31 light is discussed in \S3.
Details of the surface photometry of M32 and NGC~205 are presented in \S4 and
\S5, respectively.  A comparison of the observations to numerical simulations
is presented in \S6, and implications for the evolution of M32 are discussed
in \S7.  The main points of the paper are summarized in \S8.

\section{Observations and Basic Data Processing}
\label{optdata}

The observations were carried out over the course of four nights in
1992~October/November using the Kitt Peak National Observatory
0.9/0.6-m (primary mirror/corrector) Burrell Schmidt telescope with a
Tektronix ST2KA $2048\times2048$ CCD.  Each CCD frame has a field of
view of $68^{\min}\times68^{\min}$ and is slightly vignetted at the
corners.  The pixel scale is $2\farcs03$ and the typical FWHM of
stellar images is in the range of 2--5~pixels ($4^{\sec}-10^{\sec}$)
due to seeing and, in the worst cases, poor focus.  The data set
consists of $22\times10$-min exposures in the $B$ band and
$35\times10$-min exposures in the $I$ band.

After overscan/bias subtraction, trimming, and flat-fielding, each
individual CCD frame is geometrically transformed onto a
distortion-free astrometric system defined by the HST Guide Star
Catalog.  The images in each band are then flux calibrated to a common
photometric system, corrected for temporal variations in the night-sky
brightness, and mosaiced into a composite image.  The reader is
referred to \citet{paper0} for the details of the mosaicing technique
and least-squares method of transparency and sky-brightness
corrections.

The final $B$- and $I$-band mosaic images each cover a
$\sim$$1.7^\circ\times5^\circ$ region centered on M31.  Though both
mosaics cover all of M32, only the $B$-band mosaic covers all of
NGC~205.  The $I$-band coverage is limited to the SE half of the
galaxy.  Due to varying degrees of frame overlap, the effective
exposure time is not uniform across the entire field of view; typical
effective exposure times are 20 min in $B$ and 40 min in $I$.

\section{Removal of M31's Disk Light}

The projected distance between M31 and its two nearest satellites is
small enough that there is significant overlap in their light
distributions.  At the location of M32's center, for example, M31's
disk light accounts for $\sim$$12\%$ of the background in both $B$
and $I$ bands.  In addition, its steeply sloped contribution varies from
$\sim$$5\%-20\%$ between M32's SE and NW extremities ($r=300^{\sec}$).  An
investigation of the satellites' global properties requires a careful
treatment of this contamination.  Previous attempts to remove the M31
light contribution from the satellite profiles have relied on a simple
plane or a low-order polynomial fit to the {\it local\/} background
\citep{kent87,peletier93}.  While this approach is successful in modeling the
smooth contribution of M31's disk, it is not as effective at removing
disk features such as spiral arms.  The advantage of using a large-format CCD
mosaic image is that it allows for the {\it global\/} modeling and
subtraction of M31's disk light.

\begin{figure}
\begin{center}
\epsscale{1.0}
\plotone{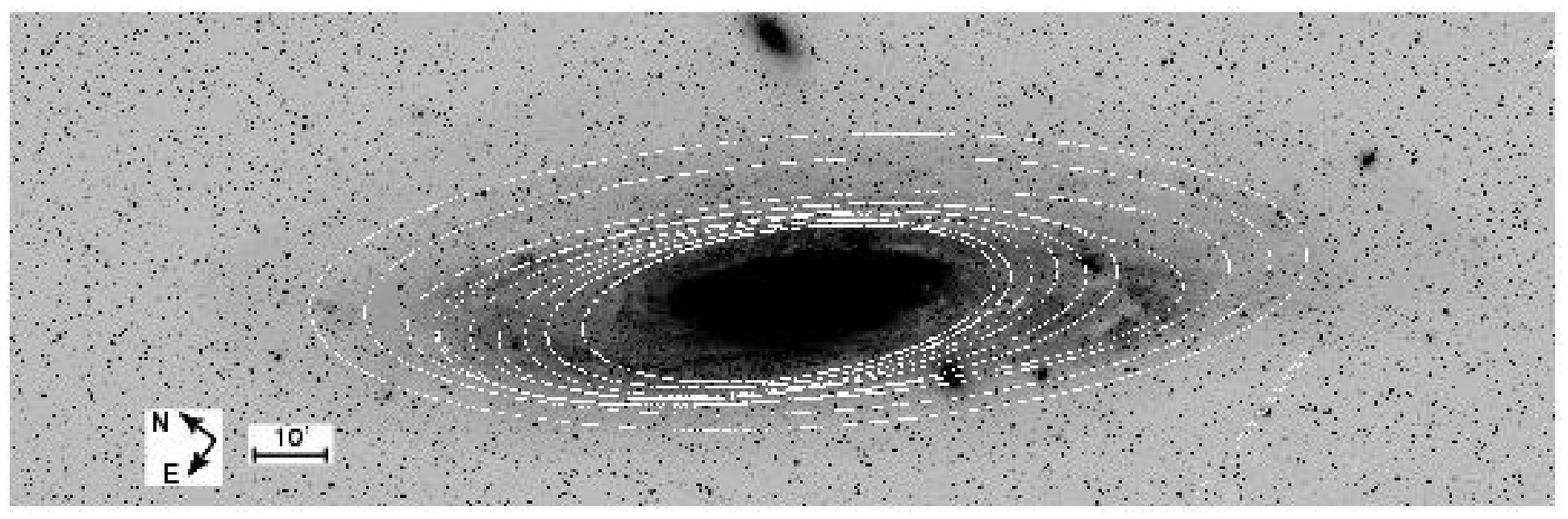}
\caption{Grayscale representation of a $B$-band $1.3^{\circ}\times
3.4^{\circ}$ image of the M31 sub-group, covering M31, M32 and
NGC~205.  The best-fit elliptical isophotes of M31 over the semi-major
axis range $30^{\min}<r<70^{\min}$ are overlaid to illustrate the
overlap with M32.
\label{fig0}}
\end{center}
\end{figure}

\begin{figure}
\begin{center}
\epsscale{1.0}
\plotone{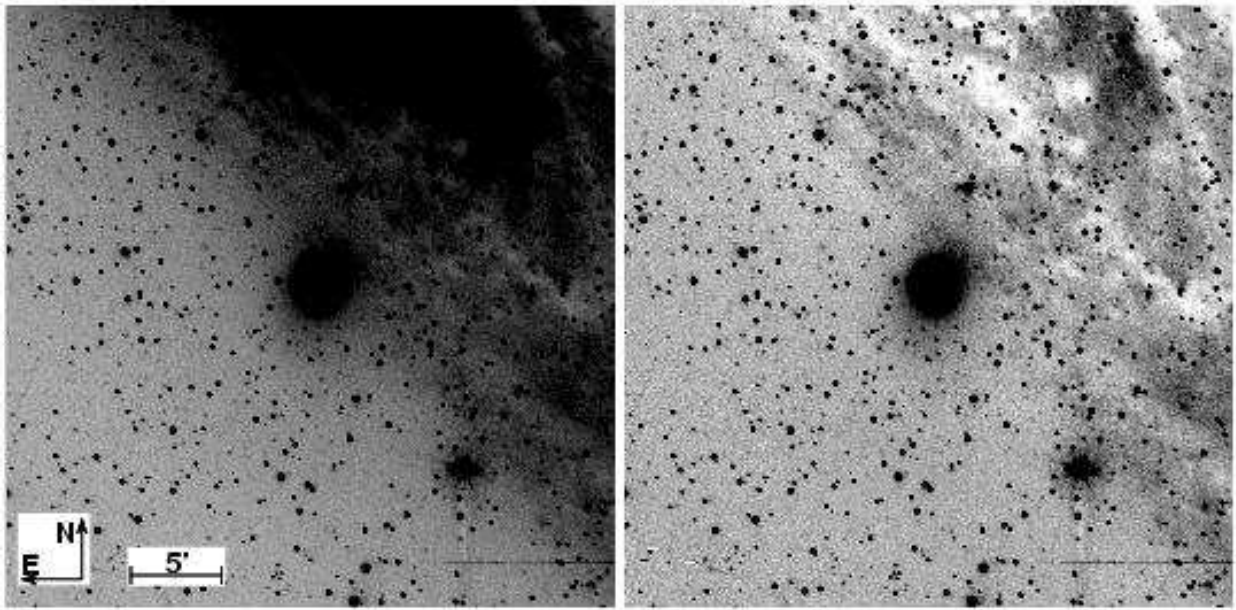}
\caption{Grayscale representations of $B$-band images centered on M32
covering $34^{\min}\times 34^{\min}$ with ({\it left}) and without
({\it right\/}) M31's disk light contribution.  Note the steep
gradient in the background across M32 caused by the inclined disk of
M31 ({\it left\/}) and the residual fine-scale structure (dust lanes,
spiral arms, etc.)  even after subtraction ({\it right\/}).
\label{fig1}}
\end{center}
\end{figure}

For the purpose of modeling the M31 disk only, the original $B$- and $I$-band
CCD mosaic images are median-filtered using a $30^{\sec}\times30^{\sec}$
window.  The resulting image is largely free of foreground Galactic stars and
compact M31 disk features.  The implementation of a two-dimensional
exponential disk plus de~Vaucouleurs bulge model reveals strong departures
from global symmetry in the form of large-scale disk warps and spiral arms.
By contrast, the more empirical approach of modeling M31 annulus by annulus
with elliptical isophotes better reproduces its global light distribution: it
provides enough flexibility to fit large- and intermediate-scale structures
on scales larger than the angular extent of the two satellites.  A series of
ellipses is fit to the M31 disk isophotes by applying the IRAF/STSDAS task
ELLIPSE to the star-removed, median-filtered images in each of the $B$ and
$I$ bands.  Figure~\ref{fig0} shows only the best-fit ellipses in the
semi-major axis range $30^{\min}<r<70^{\min}$ overlaid on the original
(i.e.,~unfiltered) $B$-band CCD mosaic image; the {\it full\/} M31 fit
extends to a semi-major axis length $r=138^{\min}$ and even overlaps with
NGC~205.  Compared to previous attempts to remove M31's disk light, the
ellipse-based coordinate system is a more natural choice for modeling the
spiral arm structure which is often sharp in the radial dimension, but
extended in the azimuthal dimension.  The best-fit ellipse models are
subtracted from the original mosaic images to create residual images
containing the satellites, largely free of M31 disk light.  Figure~\ref{fig1}
shows $B$-band images of M32, before and after M31 subtraction, emphasizing
the importance of careful subtraction.  Though the majority of M31's disk
light is well-subtracted in the latter image, fine-scale residual structure
such as dust lanes and star-forming knots in the spiral arms are still
evident.  This residual fine-scale structure is a potential source of
systematic error in the surface photometry of M32's faint outer regions
(see \S4.4).

\section{M32}
Due to its small projected separation of only 5.5~kpc from the large spiral
galaxy M31, M32 is a good test case for investigating tidal effects on
satellite galaxies.  The high surface brightness inner isophotes of M32 are
nearly circular and are well characterized by an $r^{1/4}$
law $\mu$ profile.  The inner brightness distribution provides a
rather simple (extrapolated) baseline with respect to which subtle departures
in the outer parts might be identified and measured.  These include sharp
features or ``breaks'' in the $\mu$ profile, isophotal
elongation and twists, and other signatures of tidal interaction.  As
discussed in \S1 above, such a study is relevant because: (1)~M32 is the prototype of the rare class of cE galaxies that may result from the tidal truncation of
normal E galaxies, (2)~its proximity allows for detailed observations that are currently
unavailable for any other object in its class,  
and (3)~the recently discovered stream in the outer halo
of M31 \citep{ibata+01} might be tidal debris from M32 or NGC~205.

A complicating factor in the study of M32 is the fact that it happens to be
superposed onto the face of M31.  Figure~\ref{fig2} shows M32 at two contrast
levels in each of the $I$ and $B$ bands.  The high-contrast panels on the
right reveal that a significant amount of fine-scale residual structure
remains even after our best attempts to model and subtract M31's
disk light.  This fine-scale structure is most prominent in the $B$ band on
the NW side of the M32 nucleus, toward the bright inner disk of M31; it is
probably associated with dust lanes and star-forming regions.  In \S4.4,
tests are carried out to characterize the effect of these residual
contaminating M31 disk features on measurements of the faint outer isophotes
of M32.

\subsection{Surface Brightness Profile}
Surface photometry is carried out using standard ellipse-fitting techniques with
the IRAF task ELLIPSE, independently in $B$ and $I$ bands.  Measurements are made out to a
semi-major axis length of $r\sim425^{\sec}$ (1.6~kpc), which corresponds to
a limiting surface brightness level of
$[\mu_B,~\mu_I]=[27,~25]$~mag~arcsec$^{-2}$.  Ellipse fits are performed in
three ways: on the entire galaxy, on the NW half only, and on the SE half
only.  Unless otherwise noted, the measurements of M32's surface brightness ($\mu$), isophotal
ellipticity ($\epsilon$), and isophotal position angle ($\phi$ or
$\phi^{\prime}$)\footnote{The position angle $\phi$ is defined following the
usual observational convention: anticlockwise from N (i.e.,~N through E).
For the numerical simulations, however, the position angle ($\phi$ in
Paper~I) is defined with respect to the satellite$\rightarrow$parent line
increasing towards the satellite's projected direction of motion, this being
the natural coordinate system for the simulations.  By analogy, we define the
quantity, $\rm\phi_{M32}^\prime=\pm(\phi_{M32}-1.1^\circ)$, where the
positive sign is adopted corresponding to a clockwise projected orbit for M32
around M31 (\S\S$\>$6.3--6.4).}
presented in the rest of this paper are based on ellipse fits to M32's SE half,
as it is least susceptible to contamination from M31's inner disk; the global
and NW-half ellipse fits are only used to test the symmetry of M32's
isophotes (\S4.4.1).  The central positions of the fitted ellipses are held
fixed at the nominal value determined from the innermost isophotes (the
obvious nucleus of M32), while their $\epsilon$ and $\phi^{\prime}$ are
allowed to vary with semi-major axis from ellipse to ellipse.  The best-fit
ellipses with semi-major axis length $100^{\sec}<r<300^{\sec}$ are overlaid
on the $B$- and $I$-band M32 images in Figure~\ref{fig2}, illustrating the
radial extent of the low surface brightness region.

Radial profiles of $\mu$, $\epsilon$, and $\phi^{\prime}$, derived from the
best-fit elliptical isophotes, are presented in Figure~\ref{fig3} in
$r^{1/4}$ (left) and log-linear coordinates (right).  The $B$- and $I$- band
profiles are all seen to be in good agreement with each other.  A comparison
to the $R$-band study of \citet{kent87}, shows that the $\mu$ profiles are
consistent out to his limiting measured isophote of $r\sim300^{\sec}$.  By contrast the $\epsilon$ and $\phi^{\prime}$ profiles are consistent only out
to $r\sim200^{\sec}$; beyond this radius, these isophotal shape parameters are frozen in Kent's study at their last fit values due to insufficient signal-to-noise.  Figure~\ref{fig4}{\it a\/} shows an
$I$-band image of M32, in contrast to an
M32 residual image (Fig.~\ref{fig4}{\it b\/}), in which the best-fit ellipse model for M32 has been subtracted.  The
smoothness of the latter residual image provides a measure, albeit
qualitative, of the goodness of the M32 ellipse fits.

A de~Vaucouleurs $r^{1/4}$ law profile is a convenient way to parameterize
M32's radial $\mu$ distribution.  Independent fits to the $B$- and $I$-band
data, over a $5.5$~mag range in $\mu$ from $10^{\sec}<r<140^{\sec}$, yield
best-fit $r^{1/4}$ law profiles with $r_{B,I}^{\rm eff}=29^{\sec}$,
$\mu_I^{\rm eff}=17.53$~mag~arcsec$^{-2}$, and $\mu_B^{\rm
eff}=19.43$~mag~arcsec$^{-2}$.  This ``standard'' fit is in general
agreement with Kent's $R$-band fit over the semi-major axis range
$15^{\sec}<r<100^{\sec}$: $r_R^{\rm eff}=32^{\sec}$ and $\mu_R^{\rm
eff}=18.79$~mag~arcsec$^{-2}$.  While there is excellent overall consistency
across $BRI$ bands, close inspection reveals systematic differences between
this ``standard'' $r^{1/4}$ law fit and M32's actual $\mu$ profile.  These
differences, as well as alternative $r^{1/4}$ law fits, will be explored
further in \S4.3; for the sake of comparison to previous analyses, the
``standard'' $r^{1/4}$ law fit is adopted in the following section.

\subsection{A de~Vaucouleurs Profile Excess: The ``Faint Diffuse Plume''
Revisited}

\begin{figure}
\begin{center}
\epsscale{1.0}
\plotone{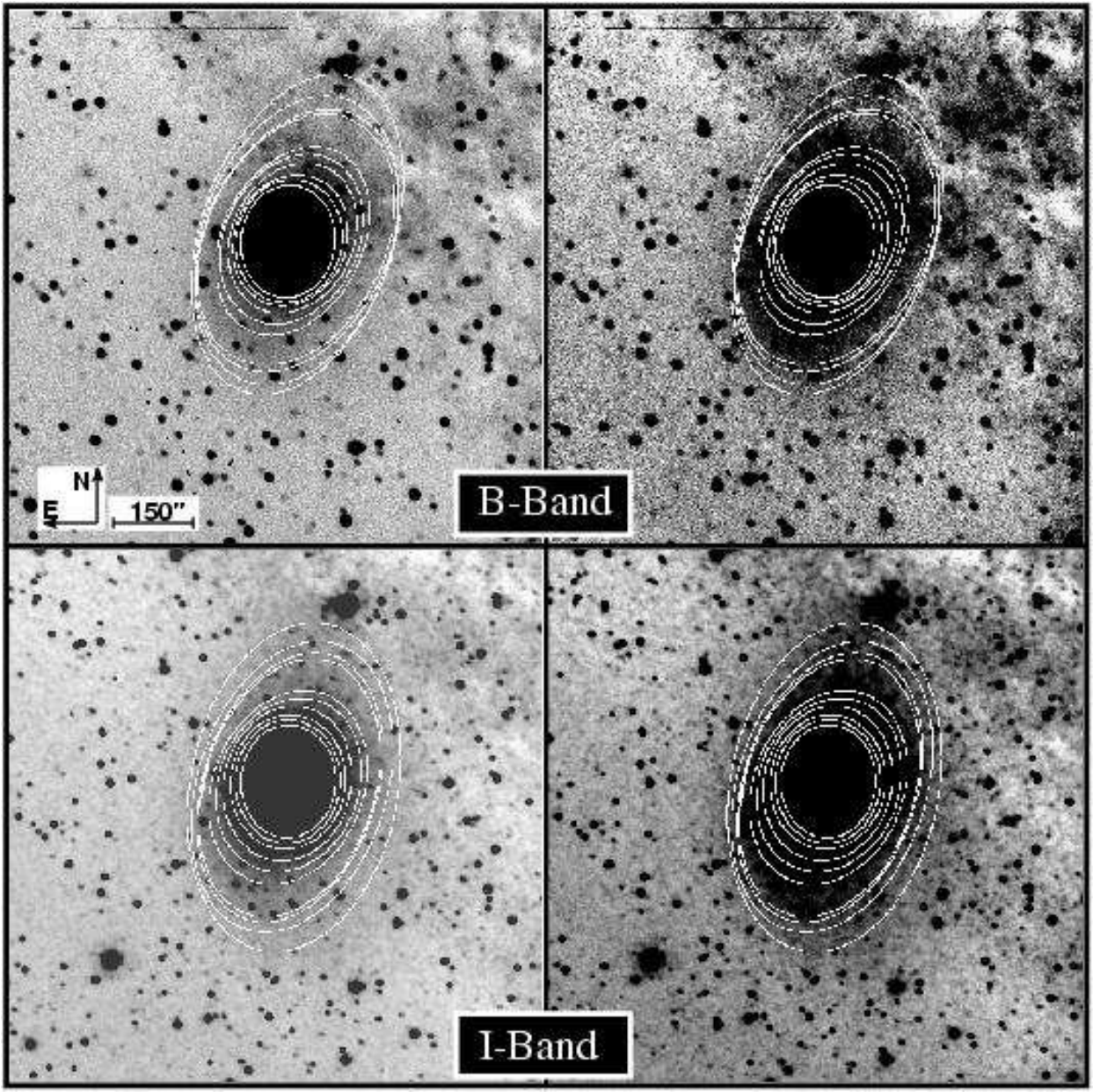}
\caption{Grayscale representations of $B$- ({\it upper\/}) and
$I$-band ({\it lower\/}) images of M32 covering
$17^{\min}\times17^{\min}$ at low ({\it left\/}) and high ({\it
right\/}) contrast, with M31's disk light subtracted.  Despite careful
attempts to model the M31 light distribution, the NW portion of
M32's outer isophotes is contaminated by residual M31 disk features.
Best-fit elliptical isophotes of M32 in the semi-major axis range
$100^{\sec}<r<300^{\sec}$ highlight the low surface brightness region
in which signatures of tidal interaction are observed.
\label{fig2}}
\end{center}
\end{figure}

\begin{figure}
\begin{center}
\epsscale{1.0} 
\plotone{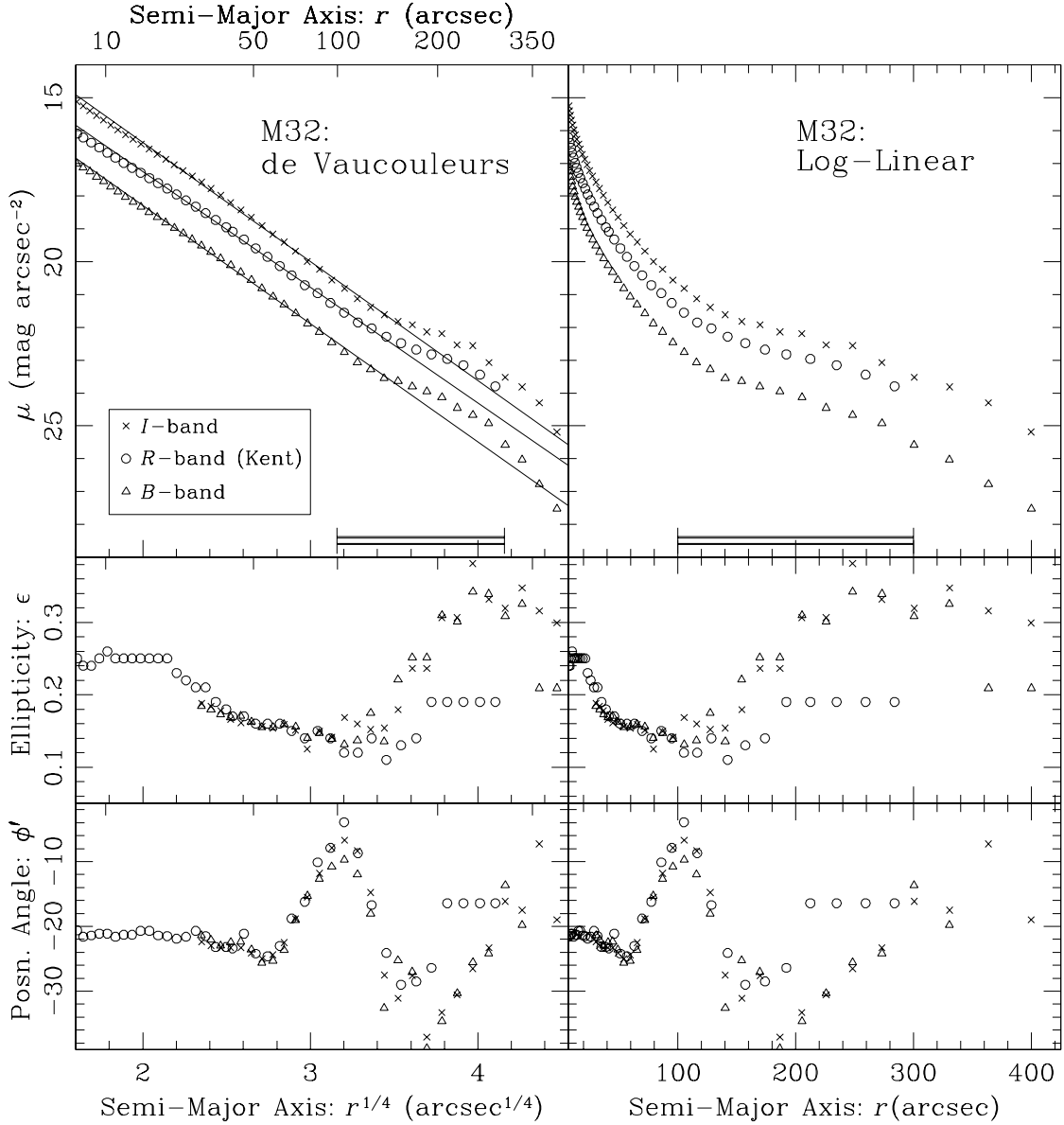}
\caption{{\it Top to bottom\/}:~~Surface brightness $\mu$, ellipticity
$\epsilon$, and position angle $\phi^{\prime}$ (measured relative to
the M32$\rightarrow$M31 vector, positive in the direction
N$\rightarrow$E [$|\phi^{\prime}_{\rm M32}|=|\phi_{\rm
M32}-1.1^{\circ}|$] of M32's isophotes versus semi-major axis length
in de~Vaucouleurs ({\it left\/}) and log-linear ({\it right\/})
coordinates in $B$ ({\it triangles\/}), $R$ ({\it circles\/}; Kent
1987), and $I$ ({\it crosses\/}).  The solid lines in the $\mu$
profile show $r^{1/4}$ law fits over a 5.5~mag range in $\mu_{I,R,B}$
over the semi-major axis range $10^{\sec}<r<140^{\sec}$ with
$r_{B,R,I}^{\rm eff}\sim 30^{\sec}$ and $\mu_{B,R,I}^{\rm eff}=19.4,
18.6$, and $17.5$~mag~arcsec$^{-2}$ (``standard'' fit in Fig~\ref{fig5}).
Note that the outer excess light feature seen in the de~Vaucouleurs
projection at $r>150^{\sec}$ is coincident with sharp shifts in the
$\epsilon$ and $\phi^{\prime}$ profiles.  The double bars marking the
range $100^{\sec}<r<300^{\sec}$ in the $\mu$ plot show the region
covered by the contours in Figure~\ref{fig2}.
\label{fig3}}
\end{center}
\end{figure}

From Figure~\ref{fig2} it is clear that although M32 appears truncated and
predominantly spherical in low-contrast images, it is surrounded by a skirt
of low surface brightness material that becomes increasingly elongated at
large radii.  This was originally detected in photographs as a ``faint
diffuse plume curved away from M31's disk'' by \citet{arp66}, and later
described by \citet{kent87} as ``an excess of light at large radii.''
Detailed characterization of this region, however, has proven to be elusive until
now.  The onset of this ``faint diffuse plume'' in the $\mu$ profile of
Figure~\ref{fig3} is marked by a clear {\it upward\/} break at
$r\sim150^{\sec}$ with respect to the ``standard'' $r^{1/4}$ law profile.
The excess is coincident with sharp shifts in $\epsilon$ and $\phi^{\prime}$
in both $B$ and $I$ bands, and is measurable to $r>300^{\sec}$ with a peak
departure of $\Delta\mu=0.5$~mag above the extrapolation of the ``standard''
fit.  
The semi-major axis range of the isophotes plotted in
Figure~\ref{fig2} ($100^{\sec}<r<300^{\sec}$) is marked with a double
line in the top panel of Figure~\ref{fig3} in order to illustrate the
region over which the excess is found.
Inspecting the relevant portion of the
image (Fig.~\ref{fig2}), it is clear why this excess feature was previously
classified by Arp as a ``diffuse plume.''  In early photographic studies, which were
sensitive to blue light, the excess region was swamped by
M31's disk structure on the NW side of M32, leading to a one-sided detection.  
Coupled with sharp changes in
$\epsilon$ and $\phi^{\prime}$ 
(Fig.~\ref{fig3}), the excess appeared to be an asymmetric
and disjoint feature alongside an otherwise well-behaved E galaxy.

Uncertainties in the surface photometry of the M32 outskirts are dominated
by systematic errors that are difficult to quantify.  Our finding of sudden
elongations and twists in the $\epsilon$ and $\phi^{\prime}$ profiles, at
radii coincident with the excess in the $\mu$ profile, indirectly indicates
that our measurements are reliable.  This is in contrast to the
\citet{kent87} study: although Kent's $\mu_R$ measurements are in general
agreement with ours, the undetermined values for
$\epsilon$ and $\phi^{\prime}$ beyond $r\gtrsim200^{\sec}$ in his study
made it difficult, at the time, to draw any firm conclusions about the
low surface brightness features in M32.

Figure~\ref{fig3} also reveals a previously undetected feature in the
form of a subtle downturn in the $\mu$ profile at $r\sim250^{\sec}$.
Although this is near the reliability limit of our data, it is seen in
both colors and is accompanied by another isophote twist in
$\phi^{\prime}$, as well as a flattening in the $\epsilon$ profile.

A note of caution may be in order here.
Ellipse parameters $\phi^{\prime}$, $\epsilon$, and, to a lesser
extent, $\mu$ are all coupled, so the mere coincidence of the profile
features does not eliminate the possibility that they are the result
of an M31 disk residual or an asymmetric feature in M32.  This possibility
can, however, be ruled out via additional tests of the background subtraction and
isophotal symmetry and color; these tests will be discussed in \S4.4.

\subsection{Evidence for an Inner Break and Depletion Zone}
Decoupling M32's tidal interaction features from its intrinsic profile
requires some a priori assumptions about its unperturbed properties.
The ``standard'' $r^{1/4}$ law fit is a good global fit to M32's
current $\mu$ profile; however, if tidal interactions have affected
its outer isophotes, alternative fits that are limited to M32's inner
regions should be more representative of its intrinsic profile.

In the left panels of Figure~\ref{fig5}, the $\mu$ profile is plotted
with the ``standard'' fit overlaid (top) along with the $\Delta\mu$
residuals of this fit (bottom).  Systematic differences are seen
between the measured profile and this best-fit $r^{1/4}$ law.  Within
the radius range $10^{\sec}<r<140^{\sec}$ (indicated by double lines)
over which the $r^{1/4}$ law is fit, these departures correspond to at
least one and possibly two additional $\mu$ profile breaks.

In the middle and right panels, the same $\mu$ profiles are shown with
``inner'' and ``extreme-inner'' $r^{1/4}$ laws that are fit to more
restricted radius ranges of $10^{\sec}<r<65^{\sec}$ and
$10^{\sec}<r<30^{\sec}$, respectively.  These alternative profiles are
shallower than the ``standard'' fit, have larger values of $r_{\rm
eff}$ ($r_{\rm inner}^{\rm eff}\sim37^{\sec}$ and $r_{\rm
extreme-inner}^{\rm eff}\sim44^{\sec}$) and fainter effective surface
brightnesses.  Table \ref{tab1} summarizes the parameters of the different
$r^{1/4}$ law fits in the different bands: the radial range over which the
data are fit, $r_{\rm eff}$, and $\mu_{\rm eff}$.

These alternative $r^{1/4}$ law fits bring a new feature to light in M32.  In
addition to the upward break at $r\sim150^{\sec}$ and its associated excess
region at large radii, there is evidence for an inner radius {\it downward}
break at $r\sim50^{\sec}$ and a ``depletion zone'' in which the surface
brightness is diminished with respect to the extrapolated $r^{1/4}$ law
profile.  Though the measured $\mu$ profile is the same in each panel, its
interpretation depends on which $r^{1/4}$ law is adopted as the
intrinsic profile.  
For instance, going from the ``standard'' to the ``extreme-inner''
fit, the residuals exhibit the general trend of a de~Vaucouleurs law
profile with an excess region at large radii to one with an
increasingly significant depletion zone at intermediate radii.  In
particular, a downward break in the $\mu$ profile can be clearly
identified in the last two residual plots at $r\sim50^{\sec}$.  This
previously unrecognized break is coincident with an inner twist in the
$\phi^{\prime}$ profile, supporting the theory that they have a
common, presumably tidal, origin.

To verify the significance of the departures discussed above, the
two-dimensional surface brightness distribution of M32 is studied.
Residual images shown in Figure~\ref{fig6}{\it b--d\/} are generated
by subtracting a de~Vaucouleurs law model of M32's light distribution
from the original M32 image.  
The models are based on the ``standard'', ``inner'' and
``extreme-inner'' $r^{1/4}$ law fits and the measured $\epsilon$ and
$\phi^{\prime}$ ellipse-fit profiles.  In each panel, a pair of
concentric circles mark the inner and outer radius limits of the
associated $r^{1/4}$ law fit.  As a contrast, the residual image of
M32's {\it actual} $\mu$ profile is shown in Figure~\ref{fig6}{\it
a\/} (same as Fig.~\ref{fig4}{\it b\/}), with all of the
aforementioned radius limits overlaid.

The images in Figure~\ref{fig6} reveal that the departures from the various
$r^{1/4}$ law profile fits are not only systematic in radius, as illustrated
in the Figure~\ref{fig5} plots, but also azimuthally symmetric.  The
systematic nature of the residuals indicate that they cannot be simply
reconciled with localized features like star formation regions.  This
global symmetry also reaffirms the investigation of the
alternative $r^{1/4}$ law profile fits.  Going from the ``standard''
to ``extreme-inner'' fit, the trend to a more prominent depletion zone and a
less prominent excess region, seen in the azimuthally averaged
residual profiles, is clearly evident in the images as well.

A priori, there is little reason to assume that one de~Vaucouleurs law fit is
more representative of the intrinsic profile than any other; however, as
will be seen in \S6, the simulations provide some useful hints.  They
indicate that generic profiles of interacting satellites all show
evidence of depletion and excess regions.  This implies that the
interpretation of M32 based on the ``inner'' and ``extreme-inner''
fits may be the most physically significant.  The implications for the
evolution and tidal interaction history of M32 will be addressed in
\S7.

\subsection{Testing the Robustness of the Measured Brightness Distribution}
Comparison of the observations to N-body simulations hinges on the
reliability of the quantities derived from the isophote fits.  It is
critical to test for potential systematic errors that may bias
the photometry.  Two approaches are taken to convince ourselves and
the reader that the measured faint features in the M32 outskirts are not
artifacts of the reduction procedure or M31 contamination.
The first is an investigation of background subtraction errors and
isophote symmetry.  The second is a color test of M32's extended
isophotes.  Discussions of these are presented below.

\subsubsection{Background Subtraction and Symmetry}

Accurate measurement of low surface brightness, extended isophotes
requires a careful characterization of the sky and spatially variable
M31 contribution.  In order to verify that features such as the upward
break in the depletion zone ($r\sim 150^{\sec}$) are not relics
of the reduction, the robustness of M32's $\mu$, $\epsilon$,
and $\phi^{\prime}$ profiles against various background errors is
tested.  The $B$- and $I$-band $\Delta\mu$ plots in Figure~\ref{fig7}
and the $\epsilon$ and $\phi^{\prime}$ plots in Figure~\ref{fig8}
illustrate the results of these tests.

The primary concern regarding the measurement of M32's outer isophotes
is the accurate removal of M31's disk contribution.  The ideal M31
disk fit minimizes residuals in the regions around M32, and not
necessarily over M31's entire disk.  A simple global
model allows for the fitting of large-scale background features; however, it also
introduces the problem that asymmetries in the disk can produce a
biased subtraction near M32.  A range of different
rejection thresholds and weighting functions for the M31 fit are
tested to minimize the impact of such asymmetries.
Ultimately, the best fit is based on the visual inspection of the
background after M31 subtraction.  In Figure~\ref{fig7}{\it a\/}, the
M32 $r^{1/4}$ law residuals, based on the ``inner'' de~Vaucouleurs
law, are compared for three images with different M31 subtractions.
The best-fit M31 subtraction is adopted in one (Case~A: triangles)
while an under-subtraction (Case~B: squares) and an over-subtraction
(Case~C: circles) of M31 features in the vicinity of M32 are adopted
in the others.  
It is interesting to note that it is only beyond $r\sim250^{\sec}$
that the residuals start to diverge.  This illustrates the robustness
of the surface photometry over the depletion zone and excess region.
The $\epsilon$ and $\phi^{\prime}$ profiles for the different M31
subtractions, shown in Figure~\ref{fig8}{\it a\/} and
Figure~\ref{fig8}{\it c\/}, are also seen to be fairly insensitive to
the M31 subtraction in both the $B$ and $I$ band.  For the remaining
de~Vaucouleurs law residual plots, the best-fit M31 subtraction is
adopted.

\begin{figure}
\begin{center}
\epsscale{1.0}
\plotone{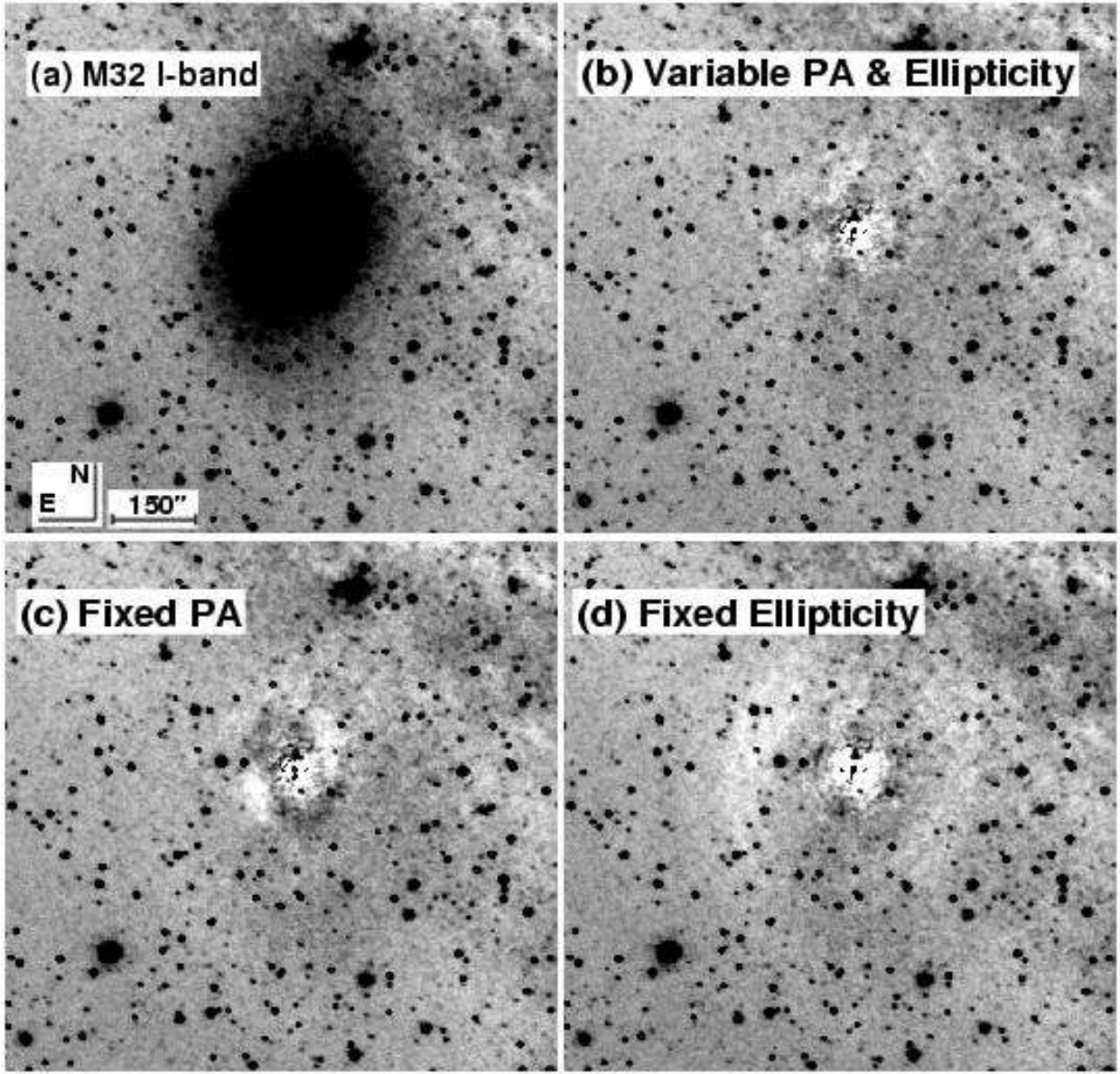}
\caption{({\it a\/})~Image of M32 in the $I$-band with M31's disk
light subtracted.~~~ ({\it b\/})~Residual image --- original M32 image
minus best ellipse fit in which the position angle $\phi^{\prime}$ and
ellipticity $\epsilon$ are allowed to vary with radius.  The contrast
level is the same as in ({\it a\/}).~~~ ({\it c\/})~Same as ({\it
b\/}), but with $\phi^{\prime}$ held constant at the inner value of
$\phi_{\rm inner}^{\prime}=-20^\circ$.~~~ ({\it d\/})~Same as ({\it b\/}),
but with $\epsilon$ held constant at the inner value of
$\epsilon_{\rm inner}=0.15$.  The orientation and scale are as in
Figure~\ref{fig2}.  Systematic features are visible in the lower
panels; these illustrate the significance of the variations of
$\phi^{\prime}$ and $\epsilon$ with radius shown in Figure~\ref{fig3}.
\label{fig4}}
\end{center}
\end{figure}

\begin{figure}
\begin{center}
\epsscale{1.0}
\plotone{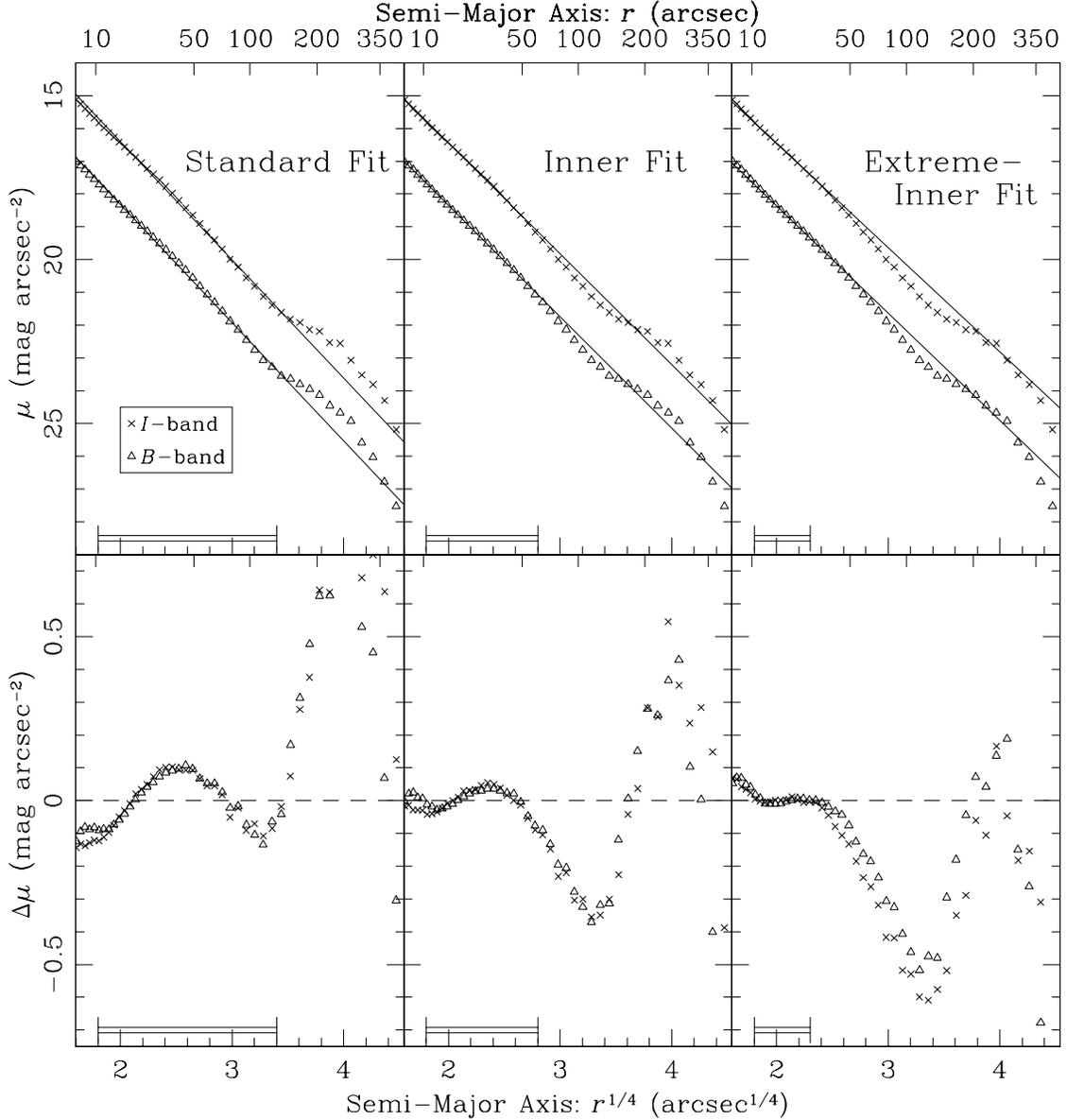}
\caption{Surface brightness $\mu$ profiles ({\it top\/}) and residuals
with respect to $r^{1/4}$ law fits $\Delta\mu$ ({\it bottom\/}) for
M32 in $B$ ({\it triangles\/}) and $I$ bands ({\it crosses\/}).
Standard ({\it left\/}), inner ({\it middle\/}), and extreme-inner ({\it
right\/}) de~Vaucouleurs law fits are shown (see \S4.3 and Table~1).
The range of radii fit is indicated by double bars along the bottom of
each panel.  Though the same $\mu$ profile is plotted in each of the
three upper panels, the differences in the fits lead to significantly
different interpretations of the residual profiles.  Clear trends are
seen from standard$\rightarrow$inner$\rightarrow$extreme-inner fit residual
profiles: the prominence of the outer excess decreases; the strength
of the depletion zone increases; and the systematic departure from the
zero line decreases over the radial range of the fit.
\label{fig5}}
\end{center}
\end{figure}

\begin{figure}
\begin{center}
\epsscale{1.0}
\plotone{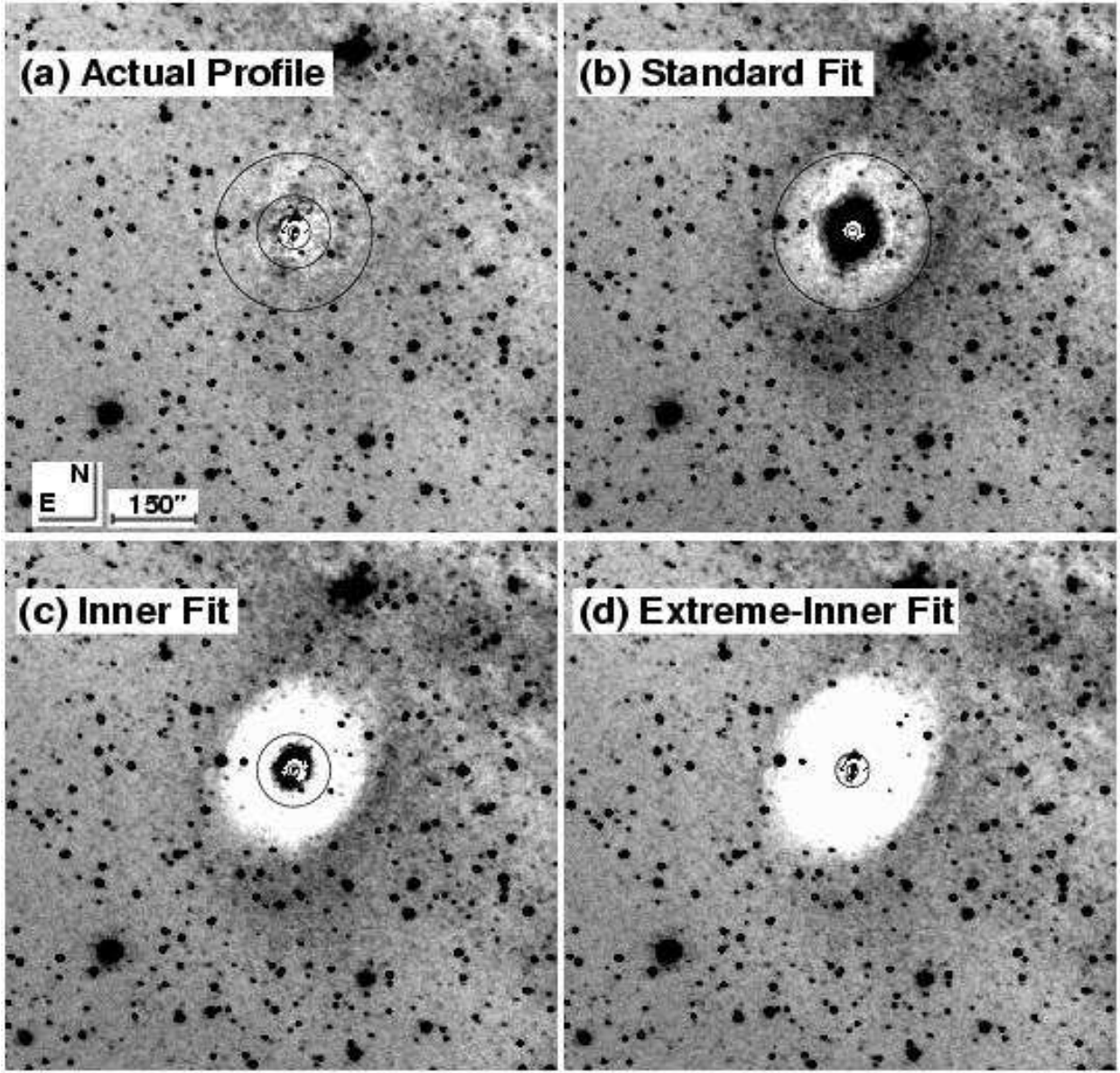}
\caption{A comparison of various M32 $I$-band residual images:~~~
({\it a\/})~Original minus best-fit ellipse model which follows M32's
{\it actual\/} surface brightness $\mu$ profile [same as
Fig.~\ref{fig4}({\it b\/})];~~~ ({\it b\/})~Original minus ellipse
model based on the ``standard'' $r^{1/4}$ law fit to the $\mu$ profile
(shown in Fig.~\ref{fig5});~~~ ({\it c\/})~Same as ({\it b\/}), but
for the ``inner'' $r^{1/4}$ law fit; and~~~ ({\it d\/})~Same as ({\it
b\/}), but for the ``extreme-inner'' $r^{1/4}$ law fit.  The orientation and
scale are as in Figure~\ref{fig4}.  The pair of concentric circles in
each of panels~({\it b--d\/}) mark the inner and outer limits of the
radial range over which the $r^{1/4}$ law is fit; all of these radii
are also plotted in ({\it a\/}).  The difference between M32 and the
``standard'' $r^{1/4}$ law fit not only varies systematically with
radius over the range of the fit (Fig.~\ref{fig5}) but is also seen to
be symmetric in azimuth ({\it b\/}).  Note, the azimuthally-symmetric
depletion zone becomes more prominent and the outer excess becomes
less prominent from ({\it b\/}$\rightarrow${\it c\/}$\rightarrow${\it
d\/}).
\label{fig6}}
\end{center}
\end{figure}

\begin{figure}
\begin{center}
\epsscale{1.0}
\plotone{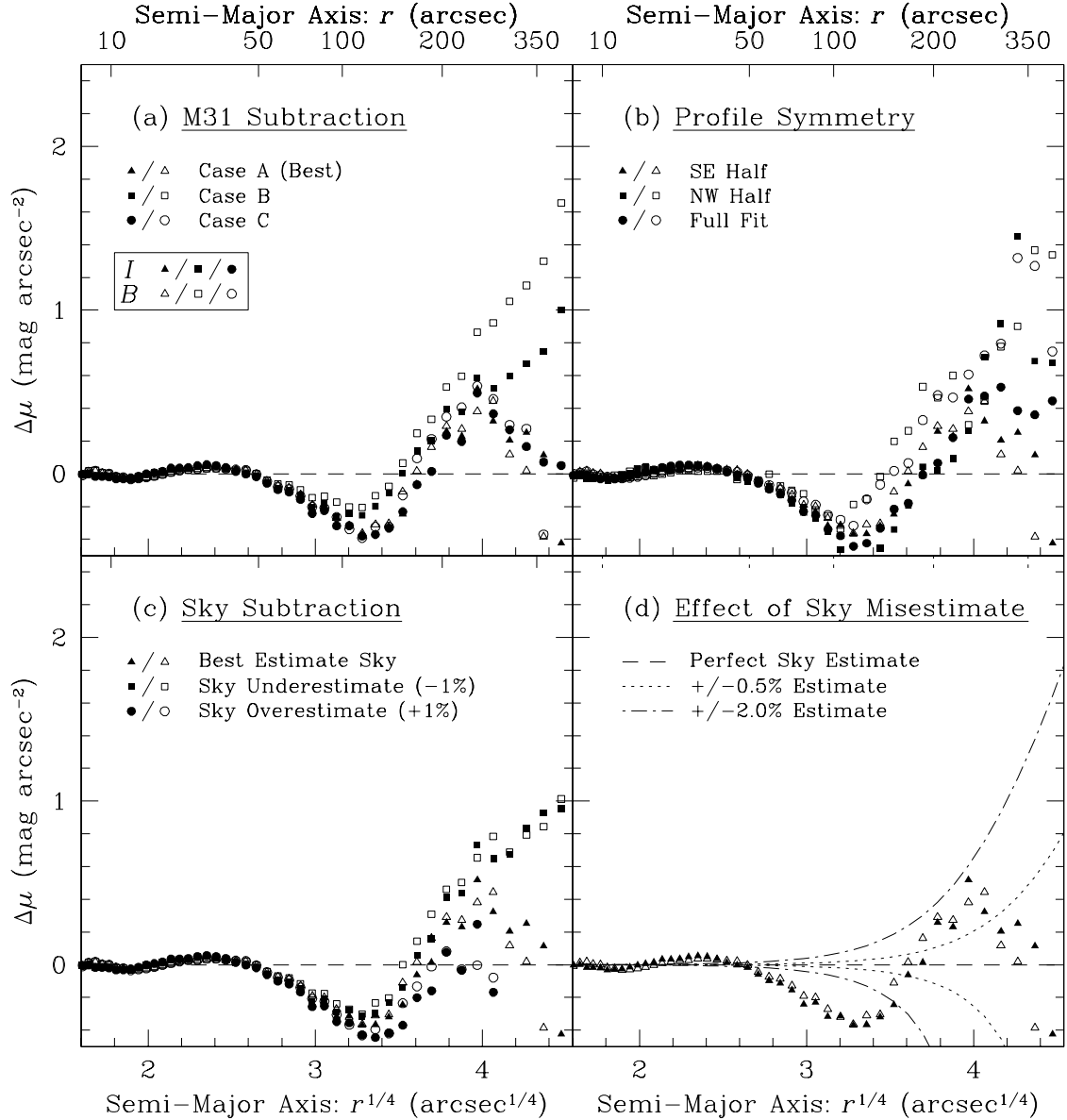}
\caption{Surface brightness residuals $\Delta\mu$ --- original minus
``inner'' fit $r^{1/4}$ law --- for M32 in $B$ ({\it open symbols\/})
and $I$ bands ({\it filled symbols\/}).  Various tests of the
robustness of M32's surface brightness profile are illustrated.~~~
({\it a\/})~Different degrees of removal of M31 disk light: best
subtraction (Case~A; {\it triangles\/}), extreme under-subtraction
(Case~B; {\it squares\/}), and extreme over-subtraction (Case~C; {\it
squares\/}) in the general vicinity of M32.~~~ ({\it b\/})~Ellipse
fits to different parts of M32: SE half of major axis towards
M31's outer disk ({\it triangles\/}), NW half of major axis
towards M31's inner disk ({\it squares\/}), and the entire galaxy
({\it circles\/}).~~~ ({\it c\/})~Different degrees of removal of a
constant sky background: best subtraction ({\it triangles\/}), 1\%
under-subtraction ({\it squares\/}), and 1\% over-subtraction ({\it
squares\/}).~~~ ({\it d\/})~Expected effect of sky subtraction error
on a galaxy with an intrinsic $r^{1/4}$ law brightness profile:
perfect sky estimate ({\it dashed\/}), $\pm0.5\%$ error ({\it
dotted\/}), and $\pm2\%$ error ({\it dot-dashed\/}).  The best
estimate of M32's residual profile ({\it triangles\/}) shows a
distinct shape that cannot simply be the result of sky subtraction
error on a galaxy with a $r^{1/4}$ law profile.  The consistency of
the M32 residuals across $B$ and $I$ bands for these various tests
({\it a--c\/}) for $r\lesssim250^{\sec}$ demonstrates the robustness
of the surface photometry and the significance of the depletion zone
and the onset of the excess region (upward break).
\label{fig7}}
\end{center}
\end{figure}

\begin{figure}
\begin{center}
\epsscale{1.0}
\plotone{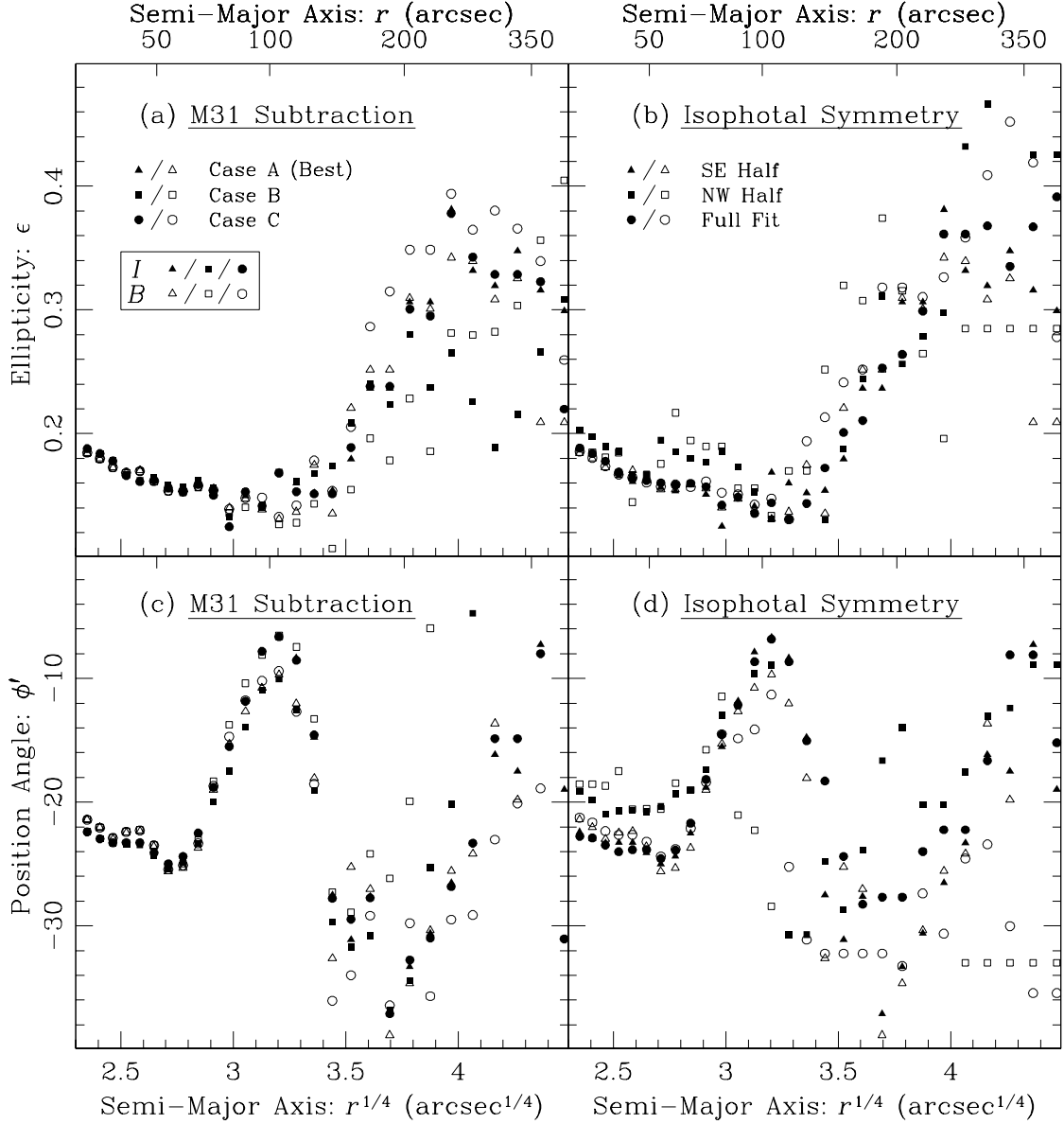}
\caption{Tests of the robustness of the isophotal shape/orientation
parameters.~~~ ({\it a, b\/})~Same as Figure~\ref{fig7}({\it a, b\/})
respectively for isophotal ellipticity $\epsilon$.~~~ ({\it c,
d\/})~Same as Figure~\ref{fig7}({\it a, b\/}) respectively for
isophotal position angle $\phi^{\prime}$.  The $\epsilon$ and
$\phi^{\prime}$ profiles are consistent between $B$ and $I$ bands,
insensitive to details of M31 disk light subtraction, and symmetric
about M32's center.  The only exception is the $B$-band
$\phi^{\prime}$ profile derived from the fit to the NW half of
M32 (indicated in (d) with {\it open squares\/}) that is likely
affected by M31 disk residual features.
\label{fig8}}
\end{center}
\end{figure}

The second concern is the careful treatment of M31's residuals around
M32, after the best-fit M31 subtraction has been performed.
Although the M32 isophote fit allows for the spatial filtering of
background sources, it is difficult to filter non-uniform variations
that, due to the steep slope in M31's disk, systematically increase in
magnitude from one side of the galaxy to the other.  The NW
side of M32 tends to suffer from more severe contamination than the
SE side, even in the M31 subtracted image.  The impact of this
variable contamination on the M32 isophote fits is tested by dividing
the galaxy along its minor axis into halves --- toward and away from
the nucleus of M31 --- and performing independent ellipse fits to both
halves, as well as to the whole.  Figure~\ref{fig7}{\it b\/} shows the
de~Vaucouleurs law $\Delta\mu$ profiles that result from the ellipse
fit to the SE half (triangles), the NW half (squares)
and the entire body of M32 (circles).  Out to a distance of
$r\sim250^{\sec}$ the scatter in this plot is low, indicating that the
$\mu$ profiles of the different fit regions are in good agreement.
Beyond this radius, the M31 residuals on the NW side start to
visibly affect the M32 surface brightness measurement.  In
Figure~\ref{fig8}{\it b\/} and Figure~\ref{fig8}{\it d\/}, similar
results are found for the $\epsilon$ and $\phi^{\prime}$ profiles,
with one exception.  The NW half $B$-band ellipse fit produces
$\epsilon$ and $\phi^{\prime}$ profiles that are noisier than the
others, and fixed beyond $\sim$$250^{\sec}$.  The $I$-band fits of this
same region, however, are in agreement with those of the SE
half, indicating that the true profile is symmetric and
that the departures seen in the $B$-band are most likely due to M31 disk
residuals.  Further evidence reinforcing the symmetry of $\epsilon$
and $\phi^{\prime}$ is seen in the final two panels of
Figure~\ref{fig4}.  In these residual images, the best-fit ellipse
model for M32 is modified to have constant $\phi^{\prime}$
(Fig.~\ref{fig4}{\it c\/}), and constant $\epsilon$ (Fig.~\ref{fig4}{\it
d\/}), with both held at their inner radius values.  The systematic
features visible in these images in comparison to the best-fit
residual image (Fig.~\ref{fig4}{\it b\/}) illustrates the significance and symmetry of the
$\epsilon$ and $\phi^{\prime}$ variations with radius.  Based on these
tests, the cleaner SE half of the galaxy is ultimately adopted
for the best-fit $\mu$, $\epsilon$ and $\phi^{\prime}$ profiles.

The final concern relates to the accurate determination of the sky,
since the outer isophotes of M32 have surface brightnesses that are a
small fraction of the sky level ($\sim$$1$--$2\%$ at $r\sim300^{\sec}$).  
Regions well away from M32 that appear to be clean of contaminating
spiral arms or dust lanes, are used to estimate the sky.  To investigate potential
systematic errors that this may introduce, the
robustness of the $\mu$ profiles is tested in the final two panels of
Figure~\ref{fig7}.
In Figure~\ref{fig7}{\it c\/}, residuals computed with our best
estimate for the sky background (triangles) are compared to those that
would result from a misestimate of the sky (squares, circles).  This
extreme example shows that even a large $\pm1$\% sky error cannot
account for the low surface brightness depletion and excess features.
This point is reinforced in Figure~\ref{fig7}{\it d\/}, in
which M32 residuals for the best-case sky subtraction (triangles) are
plotted against curves that represent the expected residuals for a
theoretical de~Vaucouleurs law profile with various degrees of sky
subtraction error.  The predictable effect of sky errors does not
match the M32 residuals, indicating that sky misestimates are not
responsible for the low surface brightness features.

Together, Figure~\ref{fig7} and Figure~\ref{fig8} show the reliability
of the $\mu$, $\epsilon$ and $\phi^{\prime}$ profiles out to at least
$r\sim250^{\sec}$ and demonstrate that the profile features associated
with the depletion zone and excess region cannot be reconciled with
background subtraction errors.  To further illustrate this point, a
color comparison of the residuals is investigated.

\subsubsection{Color Comparison of Extended Isophotes}

\begin{figure}
\begin{center}
\epsscale{1.0}
\plotone{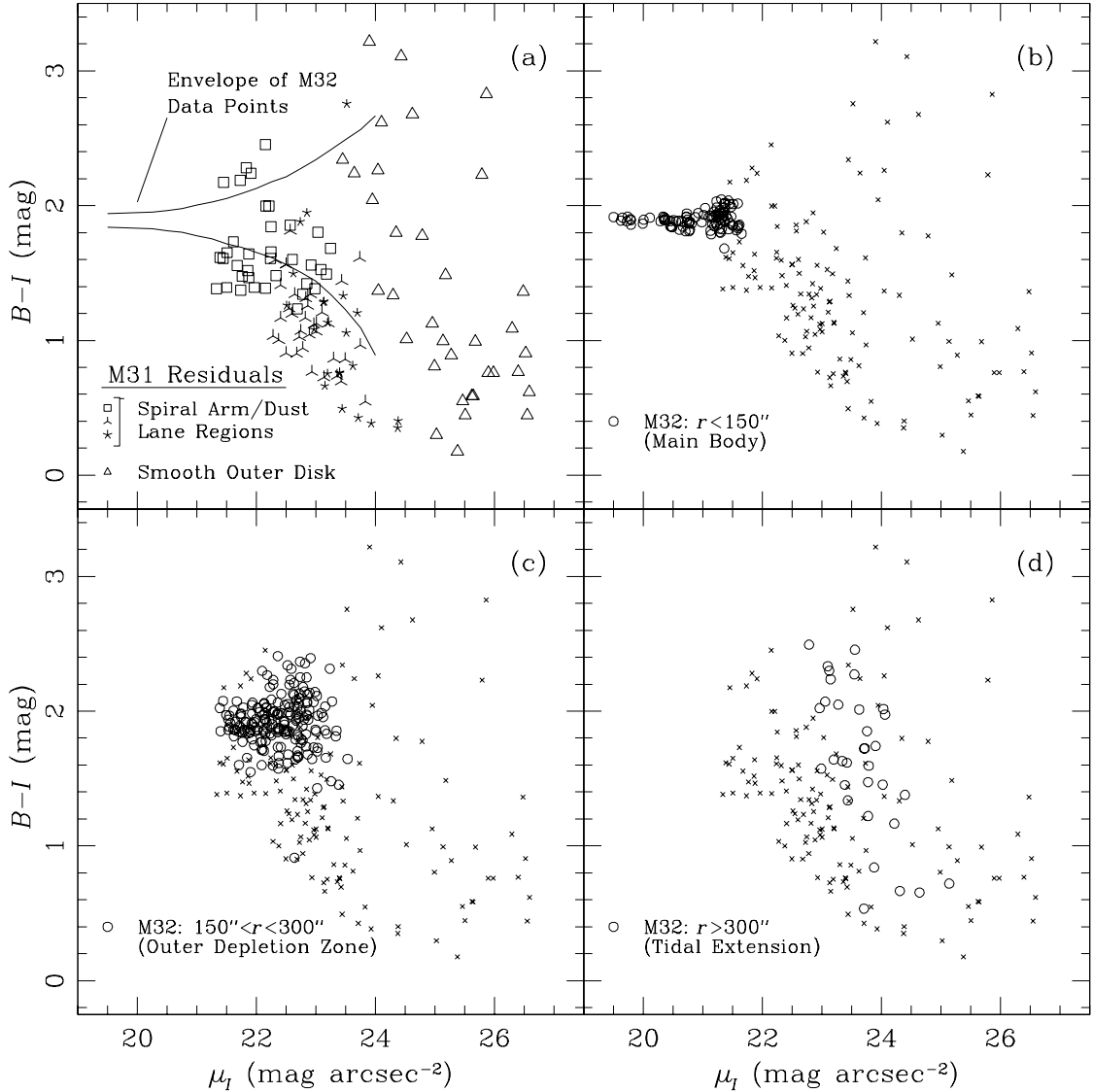}
\caption{Color versus surface brightness for the area enclosed by the
elliptical isophotes fit to M32 and areas in the surrounding field.~~~
({\it a\/})~Regions sampling M31 disk residual features in four~areas
well away from M32 ({\it stars, crosses, squares, triangles\/})
plotted against the approximate envelope of measured data points
within the area of the M32 isophotes ({\it solid lines\/}).~~~
Measurements are made for various subregions within the M32 isophotal
ellipses ({\it open circles\/}) that are then grouped into radial
bins: ({\it b\/})~$r<150^{\sec}$, ({\it
c\/})~$150^{\sec}<r<300^{\sec}$, and ({\it d\/})~$r>300^{\sec}$.
Surrounding field measurements are shown as {\it small crosses\/}
[same data points as in ({\it a\/})].  The various M32 subregions from
the bright center to the faint outer isophotes ($r\sim300^{\sec}$)
form a well-defined horizontal locus, in contrast to M31 residual disk
features in the wider field which tend to be bluer.  This suggests
that the isophotes in the range $150^{\sec} <r<300^{\sec}$ (over which
tidal signatures are observed) are indeed associated with M32 with
relatively little contamination by residual M31 disk features.
\label{fig9}}
\end{center}
\end{figure}

It is impossible to remove every small-scale M31 disk
feature; therefore, the possibility
that the extended isophotes of M32 are dominated by a chance
superposition of these residual features is investigated.
Although background contamination would
generally produce asymmetric features, a color comparison of
the extended isophotes, to those of the sky and the M31 disk residuals
provides a useful complementary test.  Using the $B$- and $I$-band
images, a color-index map is made of M32 and its surroundings.
In Figure~\ref{fig9}, different sections of this map are sampled in
$10^{\sec}\times10^{\sec}$ boxes and plotted on a $B-I$ versus $\mu_I$
diagram.  Finely sampling each isophote of M32, from the core ($r<10^{\sec}$) 
to the most extended isophotes ($r>300^{\sec}$), the mean value of the
M32 color index is determined to be $B-I=1.9$ with relatively small scatter for
regions within $r<150^{\sec}$.  An envelope enclosing the
locus of points representative of M32 is plotted in
Figure~\ref{fig9}{\it a\/}.  The points in this panel sample three
distinct M31 residual regions (stars, crosses, squares) as well as
smooth patches of uncontaminated sky (triangles).

In the three remaining panels (Figs.~\ref{fig9}{\it
b\/}--{\it d\/}) this background sample (small crosses)
is plotted against points (circles) that represent three subregions
within M32's measured isophotes: the main body ($r<150^{\sec}$); the
depletion and excess regions ($150^{\sec}<r<300^{\sec}$); and the most
extended measured isophotes ($r>300^{\sec}$).  The consistency of the
$B-I$ color index in regions out to $r\sim300^{\sec}$ compared to the
colors of the field samples, supports the claim that the excess light
is truly associated with M32 and not with M31 or sky variations.  Even
beyond $r>300^{\sec}$, the color spread around the M32 mean color is
relatively small down to $\mu_I^{\rm lim}=23.5$~mag~arcsec$^{-2}$.

How does the broadband color of the tidal debris in M32's outskirts compare
to that of the stellar stream found by \citet{ibata+01} in the M31 halo?
The $B-I$ color is about 2 for M32's outer tidal excess (Figs.~\ref{fig3} and
\ref{fig9}).  Since $V-I\approx0.5(B-I)$ across a wide range of stellar types
\citep{bvdb01}, the $V-I$ color index for M32 is estimated to be about 1.  At
face value, the luminous giants near the tip of the red giant branch detected
by Ibata et~al.\ are redder than this with $V-I$ spanning the range 1--3 (see
Fig.~2{\it c\/} of their paper).  However, it should be noted that the {\it
integrated\/} $V-I$ color of an old stellar population is comparable to the
color of stars near the base of the red giant branch (cf.~Guhathakurta
et~al.\ 1998), which corresponds to $V-I\sim1$ for the stream.  Thus, the
expected $V-I$ color of the integrated stream agrees quite well with that of
M32's outer tidal material.

\section{NGC~205}
The observational data for NGC~205 are similar to those presented for
M32 above; the same surface photometry and analysis methods are
used.  The situation is somewhat easier for NGC~205 with regard to
contamination by M31 disk light, so extensive tests of the robustness
of the results are not carried out for this galaxy.  The
interpretation of the NGC~205 data, however, are far more complicated
for a few reasons: (1)~the satellite is significantly flattened, making
it difficult to draw conclusions from our simulations of spherical,
non-rotating satellites; (2)~the inner $\mu$ profile is complicated,
intermediate in Sersic index $n$ between an exponential law ($n=1$)
and a de~Vaucouleurs law ($n=4$) but not a good fit to any $n$ value,
making it difficult to estimate the intrinsic profile of NGC~205; and
(3)~the intrinsic brightness distribution is patchy in places, with hints of
dust lanes and star forming regions, complicating the search
for subtle departures from isophotal symmetry due to tidal effects.

\begin{figure}

\begin{center}
\epsscale{1.0} 
\plotone{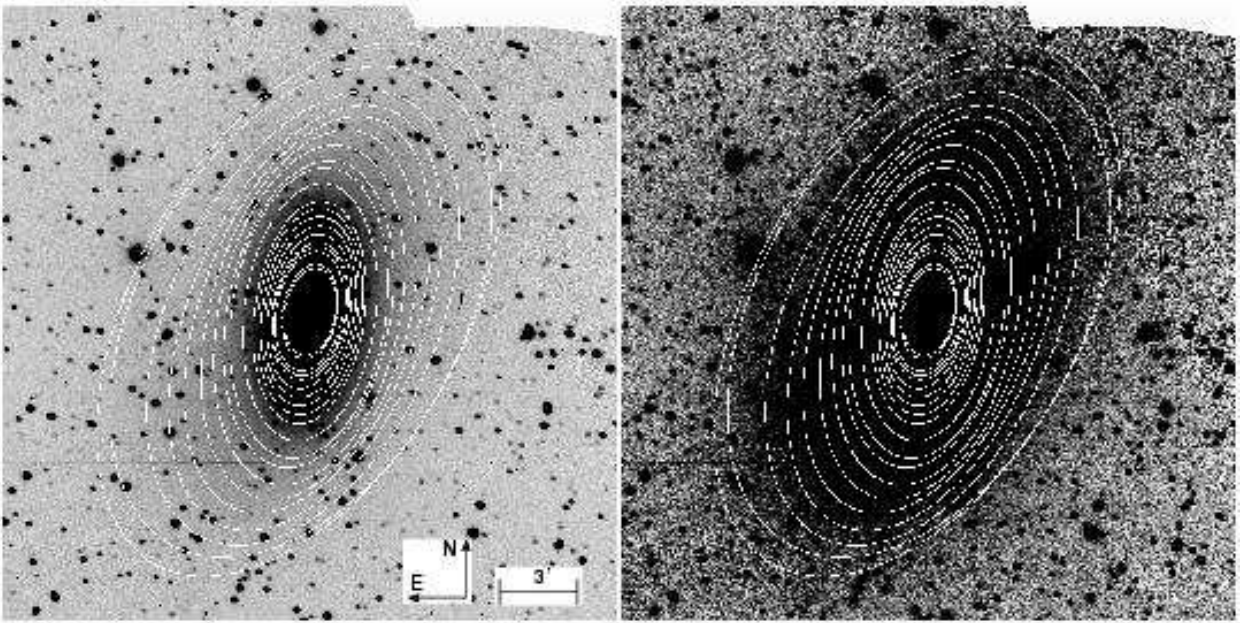}
\caption{Grayscale representations of $B$-band images of NGC~205
covering $24^{\min}\times24^{\min}$ at low ({\it left\/}) and high
({\it right\/}) contrast, with M31's disk light subtracted.  Best-fit
elliptical isophotes in the semi-major axis range
$140^{\sec}<r<660^{\sec}$ highlight the region in which pronounced
isophote twisting is observed.
\label{fig10}}
\end{center}
\end{figure}

\begin{figure}
\begin{center}
\epsscale{1.0}
\plotone{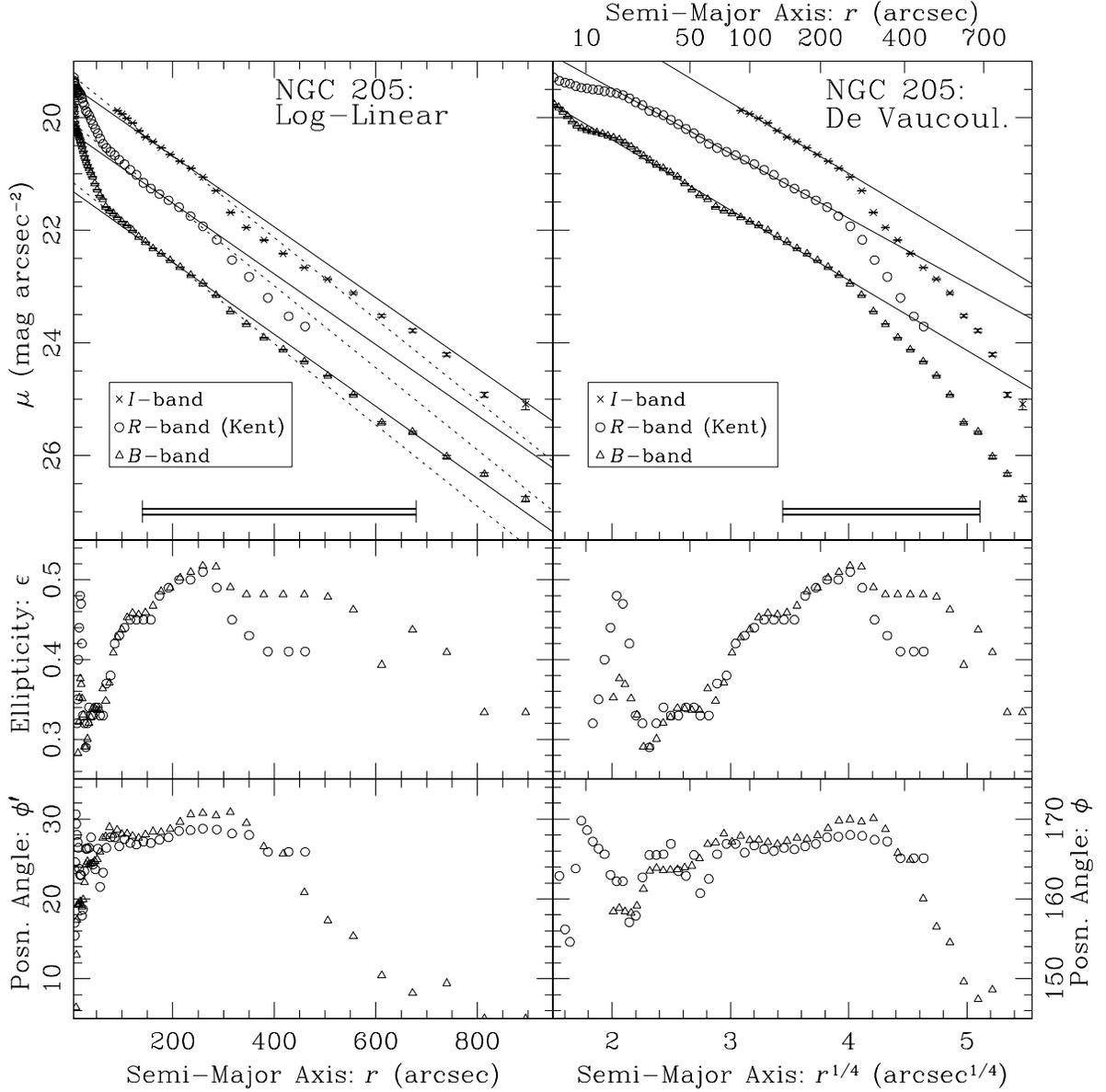}
\caption{{\it Top to bottom\/}:~~Surface brightness $\mu$, ellipticity
$\epsilon$, and position angle $\phi^{\prime}$ (measured relative to
the NGC~205$\rightarrow$M31 vector, positive in the direction
N$\rightarrow$E [$|\phi^{\prime}_{\rm NGC~205}|=|\phi_{\rm
NGC~205}-132.9^{\circ}|$]) of NGC~205's isophotes versus semi-major
axis length in log-linear ({\it left\/}) and de~Vaucouleurs ({\it
right\/}) coordinates in $B$ ({\it triangles\/}), $R$ ({\it
circles\/}; Kent 1987), and $I$ ({\it crosses\/}; only $\mu$ data due
to partial coverage of the $I$-band images).  An exponential law with
$r_{B,R,I}^{\rm exp}\sim170^{\sec}$ fits the profile over the range
$150^{\sec}<r<250^{\sec}$ ({\it solid lines\/}), while one with
$r_{B,R,I}^{\rm exp}\sim 150^{\sec}$ fits over the range
$75^{\sec}<r<250^{\sec}$ ({\it dashed lines\/}).  Note the subtle {\it
downward\/} break at $r\sim300^{\sec}$, coincident with a sharp change
in the $\epsilon$ and $\phi^{\prime}$ profiles.  The double bars
marking the range $140^{\sec}<r<680^{\sec}$ in the $\mu$ plot show the
region covered by the contours in Figure~\ref{fig10}.
\label{fig11}}
\end{center}
\end{figure}

Surface photometry is performed using the task ELLIPSE on
M31-subtracted images; however, unlike the M32 isophotes, the two
bands are not fit independently.  Due to incomplete coverage in the
$I$ band, only the $B$ band is used for the determination of the
best-fit elliptical isophotes.  The $\epsilon$ and $\phi^{\prime}$
($|\phi^{\prime}_{\rm NGC~205}|=|\phi_{\rm NGC~205}-132.9^{\circ}|$)
profiles from this fit are then used to compute the $I$-band
photometry.  Though this is not expected to have a major impact on the
resulting profiles, it does mean that the ellipticity $\epsilon$ and
position angle $\phi^{\prime}$ profiles are only fit in one band.
Surface photometry is measured out to a limiting semi-major axis
length of $r\sim720^{\sec}$ (2.7~kpc) with a limiting surface
brightness of $\mu_B^{\rm lim}=27.0$~mag~arcsec$^{-2}$.  In
Figure~\ref{fig10}, a $B$-band image of NGC~205 is shown at two
different contrast levels, with elliptical isophotes ranging from
$140^{\sec}<r<660^{\sec}$ overlaid to illustrate the
low surface brightness region in which pronounced isophote twisting is
observed.  In Figure~\ref{fig11}, the results of the best-fit
elliptical isophotes are presented as radial profiles of $\mu$,
$\epsilon$ and $\phi^{\prime}$ in log-linear and de~Vaucouleurs
coordinates.

Comparisons of the measured NGC~205 brightness profile to those of
\citet{hodge73} and \citet{kent87} show general agreement.  The data
confirm that the profile is not well fit by any one analytic profile,
but instead, is intermediate between an exponential and $r^{1/4}$ law
as noted by Kent.  Like M32, the profile fits are highly dependent on
the radius range chosen.  As is evident in the log-linear plot of
Figure~\ref{fig11}, the profile is well fit by an exponential law with
a scale length $r_{B,R,I}^{\rm exp}=150^{\sec}$, over the range
$75^{\sec}<r<250^{\sec}$ and $r_{B,R,I}^{\rm exp}=170^{\sec}$, over
the range $150^{\sec}<r<250^{\sec}$.  With respect to the
$r_{B,R,I}^{\rm exp}=150^{\sec}$ profile, a subtle downward break is
detected at $r=300^{\sec}$ in both the $B$ and $I$ bands of the
current data set, as well as in Kent's $R$-band data.  The degree of
the departure is inconsistent between the two sets.  This may be due
to increasing magnitude errors at isophotes that are approaching the
Kent brightness limit.  At radii beyond the limits of either Kent or
Hodge ($r>500^{\sec}$), the surface brightness returns to the
projected exponential profile.  Although the magnitude of this break
is subtle enough that neither Kent nor Hodge make note of it, its
coincidence with shifts in $\epsilon$ and $\phi^{\prime}$ provide
compelling evidence for its significance.

The general shape of the measured $\epsilon$ curve, which rises with
increasing radius to a maximum value of $e=0.52$ at $r=260^{\sec}$ and
then dips at larger radii, is in agreement with past results
\citep{richter_hogner63, hodge73, kent87}.  Beyond the peak at
$r=260^{\sec}$ there is some discrepancy between the curves of Hodge
and Kent.  Kent shows a sharply dropping ellipticity out to the limit
of his data at $r=460^{\sec}$, while Hodge sees only a slight dip at
$r=400^{\sec}$ followed by a continued rise to the end of his data at
$r=480^{\sec}$.  The M31 removed CCD measurements confirm a slight dip
beyond the maximum and then a gradual decline of the ellipticity out
to and beyond $480^{\sec}$.  In the inner regions, the large amplitude
ellipticity fluctuation at $r\sim20^{\sec}$, seen by both Richter \&
Hogner and Kent is also confirmed.  Contrary to Hodge's speculation
that this feature is an artifact of ``a combination of poor statistics
and systematic effects,'' our data indicate that it is significant.

The major-axis $\phi^{\prime}$ profile shows a gradual increase out to
a radius of $r=260^{\sec}$ and then a steady drop corresponding to a
$30^{\deg}$ twist out to the last measured isophote.  This is in good
agreement with Hodge, who measures continuous twisting out to the limit 
of his data.

\section{Interpretation of Observations in Light of N-body Simulations}
The previous sections focused on the detailed characterization of the
M32 and NGC~205 observations.  In the following section, these
characteristics will be compared to numerical simulations, with the hope of
determining whether they are tidally induced, and if so, using them to constrain the
satellites' orbital parameters.  A brief description of the
simulations and their analysis is presented, along with results of
their application to M32 and NGC~205.  The details of the simulations
and the general trends of satellite interaction can be found in
\citet{paper2b} (hereafter Paper~I).
 

It is worth noting that the discussion in this section is based on the
similarity in appearance between the observations and simulations, rather
than a definitive proof that tides are responsible for the observed
features.  The models are not specifically tailored to match the
intrinsic properties of M32 and NGC~205, or the precise potential of
M31.  Despite this, they provide a qualitative understanding of the
physical mechanism that drives the tidal signatures.  It has been
shown that the two observed satellites have very different structural
parameters.  The current set of simulated spherically-symmetric
satellites are well suited for the analysis of M32; however, they are
not as applicable to NGC~205, due to its flattened structure.  As
a result, the bulk of the quantitative analysis is performed for
M32, and a more conservative approach is taken for NGC~205.
Fine tuning of the models and spectroscopic observations to determine
satellite internal kinematics, both of which are in progress, will
provide leverage to further refine orbital parameters and allow for a
more comparable analysis of the two satellites in the future.

\subsection{The Simulations}
In the numerical simulations, 64,000-particle, one-component,
spherical satellites are followed for five radial oscillations as they
orbit in a fixed three-component potential, representative of the
disk, bulge, and halo of a parent galaxy.  Particle interactions are
computed using code developed by \citet{hernquist_ostriker92} and
based on the basis-function-expansion technique.  Of the five
simulated satellites, four have Plummer initial density profiles and
orbital eccentricity ranging from $0.10<e<0.88$ (Models~1--4).
The fifth has a shallower Hernquist initial profile and an
eccentricity of $e=0.88$ (Model~5).

The analysis of the simulations is performed with a parallel and
complementary approach to that of the observations.  This facilitates
direct comparison between the two. Snapshot ``images'' of each
simulated satellite are generated by projecting the satellite
particles onto a two-dimensional plane and smoothing the resulting
distribution.  The images for a range of orbital phases and viewing
angles are then analyzed with the same ellipse-fitting technique that
is used for the M32 and NGC~205 images.  The resulting trends in the
surface brightness $\mu$, ellipticity $\epsilon$ and position angle
$\phi^{\prime}$ profiles with orbital eccentricity, phase and viewing
angle, are used to guide the interpretation of the M31 satellite
observations.

\subsection{Viewing Angle}


The detection of isophote twists in the simulated satellites is a
signature of an inclined orbital plane.  By contrast, when viewed from
within the orbital plane, the fitted ellipses line up along the
direction of motion and twists are not observable.  Comparing
simulations viewed at angles of $0^{\circ}$, $30^{\circ}$,
$60^{\circ}$, and $90^{\circ}$ from the orbital plane (edge-on to
face-on), isophote twists are measurable for $i\gtrsim30^{\circ}$,
indicating that even at low inclinations, the effects of tidal
twisting are observable.  Our study shows that for $i\gtrsim30$, the
observed quantities have a negligible dependence on the viewing angle.
This simplifies the analysis, but it also limits the viewing angle
determination to ``edge-on'' vs. ``face-on''.

The observed satellites, M32 and NGC~205, exhibit varying degrees of
isophote twisting, indicating face-on viewing angles; however, in the
case of NGC~205, where the assumption of intrinsic spherical symmetry
may break down, a note of caution must be added.  An intrinsically
non-spherical satellite can exhibit isophote twists for $i=0^{\deg}$
if the satellite itself is inclined with respect to its orbital plane.
NGC~205's flattened structure may have been tidally induced; but if not,
little can be concluded about its orbital inclination.

\subsection{Orbital Eccentricity and Phase}
As defined in Paper~I, the position angle $\phi^{\prime}$ is the angle
of the satellite semi-major axis, with respect to the
satellite$\rightarrow$parent galaxy vector.  It is measured on the side of
the satellite closer to the parent galaxy, so that
$-90^{\circ}<\phi^{\prime}<90^{\circ}$.  
As will be shown in \S6.4, the probable sense of M32's projected orbit
is clockwise around M31 so this is hereby adopted for the sign
convention of $\phi^{\prime}$ for both galaxies.

For circular orbits of spherical satellites $\phi^{\prime}$($r_{\rm
break}$), the position angle of the $r_{\rm break}$ isophote, and
$d\phi^{\prime}/dr$($r_{\rm break}$), describing the isophote twist,
both have the same orientation (Johnston et~al.\ 1999a).  The fact
that neither M32 nor NGC~205 exhibits this trend between
$\phi^{\prime}$($r_{\rm break}$) and $d\phi^{\prime}/dr$($r_{\rm
break}$) reveals that these satellites are not likely to be on
circular orbits.  In addition to the relationship between
$\phi^{\prime}$($r_{\rm break}$) and $d\phi^{\prime}/dr$($r_{\rm
break}$), in the case of M32, three other profile features are
suggestive of a highly eccentric orbit: (1)~the triple break in the
$\phi^{\prime}$ profile, (2)~the ratio $r_{\rm break}/r_{\rm tide}$
and (3)~the ratio $r_{\rm break}/r_{\rm distort}$.  The diagnostics
$r_{\rm break}$ and $r_{\rm distort}$ are empirically measured radii
that characterize the $\mu$ and $\epsilon$ profiles.  Specifically,
$r_{\rm break}$ is the radius at which a sharp change is measured in
the slope of the $\mu$ profile, and $r_{\rm distort}$ is the
corresponding radius for the $\epsilon$ profile (Paper~I).  By
contrast, $r_{\rm tide}$ is an estimate for the theoretical King
tidal radius.

\begin{figure}
\begin{center}
\epsscale{1.0}
\plotone{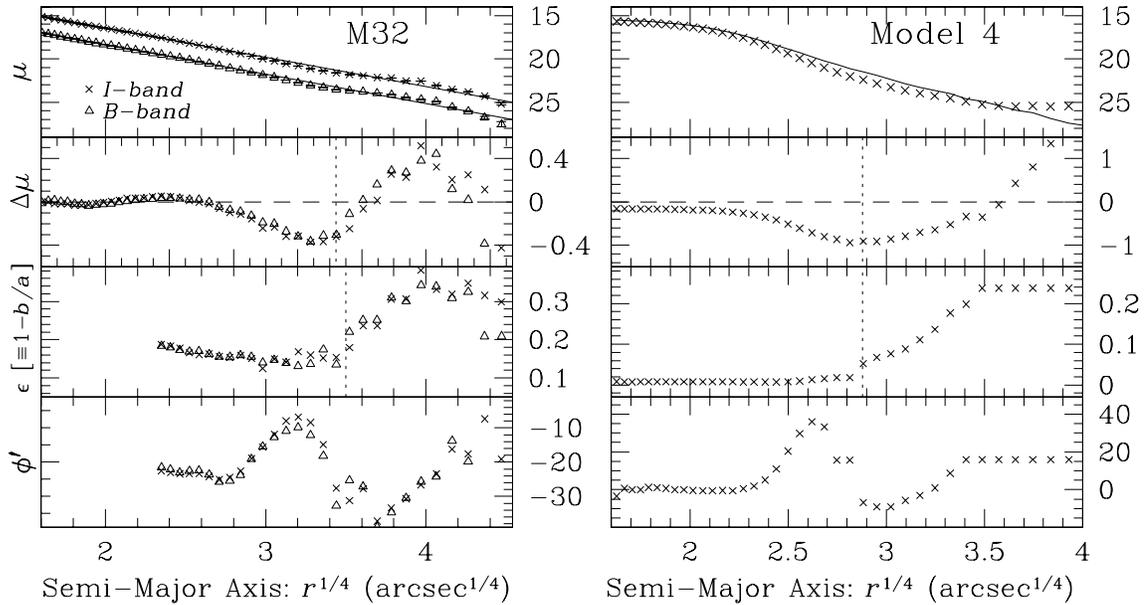}
\caption{{\it Top to bottom\/}:~~Surface brightness $\mu$, surface
brightness residual $\Delta\mu$, ellipticity $\epsilon$, and position
angle $\phi^{\prime}$ (measured relative to the
satellite$\rightarrow$parent vector) profiles of M32 ({\it left\/}) in
$B$ ({\it triangles\/}) and $I$ ({\it crosses\/}) bands and of a
snapshot of a simulated satellite with high orbital eccentricity
($e=0.88$) at an orbital phase preceding apocenter ({\it right\/}).
The residual $\Delta\mu$ is measured relative to the ``inner''
$r^{1/4}$ law fit for M32 (Fig.~\ref{fig5}) and relative to the
initial profile for the simulated satellite ({\it solid lines in top
panels\/}).  The locations of $r_{\rm break}$ and $r_{\rm distort}$
are shown as dotted vertical lines in the $\Delta\mu$ and $\epsilon$
plots, respectively.  The two sets of profiles show similar features,
including the unusual triple break in the $\phi^{\prime}$ profile,
indicating that M32 is likely to be on an eccentric orbit approaching
apocenter.
\label{fig12}}
\end{center}
\end{figure}

In Figure~\ref{fig12}, the best-fit elliptical isophote profiles for
M32 are compared to those of a simulated satellite on a highly
eccentric orbit ($e=0.88$) that is approaching apocenter (Model~4).
Striking similarities seen in the $\mu$, $\Delta\mu$, $\epsilon$, and
$\phi^{\prime}$ profiles of the observed and the simulated satellite
imply that the M32 features have tidal interaction origins.  The
$\Delta\mu$ profile is based on the ``inner'' de~Vaucouleurs law fit
for M32, and the intrinsic $\mu$ profile for the simulated satellite.
The $M32 \mu$, $\Delta\mu$ and $\epsilon$ profiles have generic shapes
that are common for many of the simulated snapshots, independent of
the satellite's orbit or phase.  The $\phi^{\prime}$ profile, however, with its
multiple twists --- each of which is coincident with either a $\mu$ or
$\epsilon$ feature ---  is more atypical.  The triple twist in
$\phi^{\prime}$, which is seen only in simulated satellites
approaching apocenter of highly eccentric orbits, provides a clue not
only about M32's orbital eccentricity, but also its orbital phase.

\begin{figure}
\begin{center}
\epsscale{1.0}
\plotone{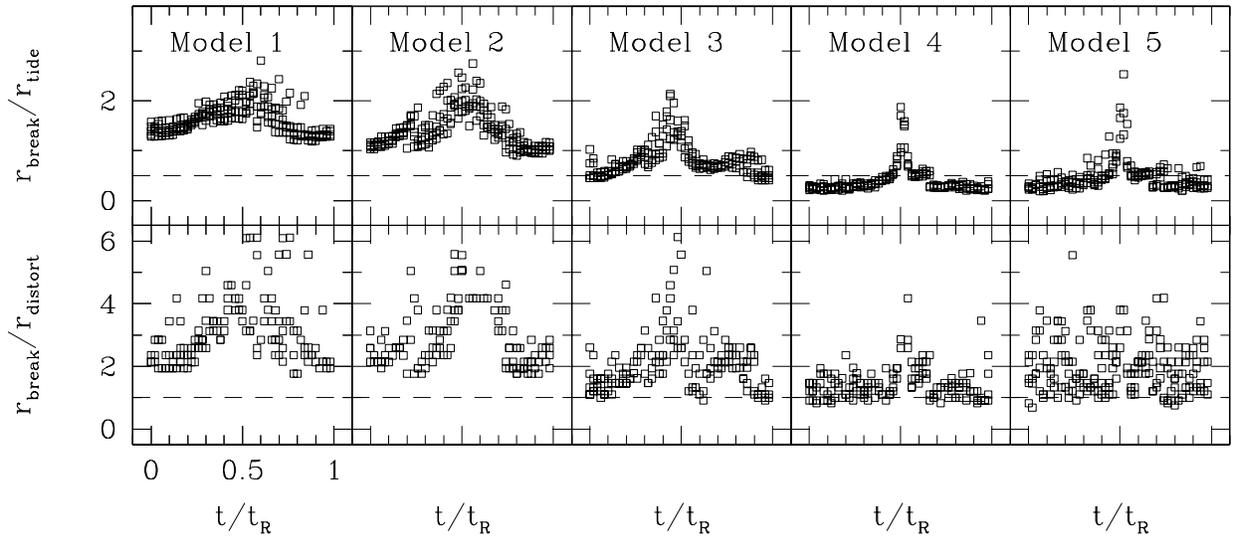}
\caption{The ratios $r_{\rm break}/r_{\rm tide}$ ({\it upper\/}) and
$r_{\rm break}/r_{\rm distort}$ ({\it lower\/}) plotted versus orbital
phase as measured in numerical simulations of tidally disrupted
satellites.  Satellite orbital eccentricities are
$e=0.10/0.29/0.67/0.88$ for Model~1 (nearly circular) through Model~4
(highly elongated), respectively.  Model~5 follows the same orbit as
Model~4, but adopts a shallower initial density profile for the
satellite than Models~1--4.  The measured ratios for M32 ({\it dashed
lines\/}) indicate that it is likely to be on an eccentric orbit
($e_{\rm M32}\ge0.5$).
\label{fig13}}
\end{center}
\end{figure}

The second signature of an eccentric orbit is $r_{\rm break}/r_{\rm
tidal}$, the ratio of the observed break in the $\mu$ profile and the
classically defined theoretical King tidal radius.  In the simplifying
case of a circular orbit, $r_{\rm tide,\,peri}$ [defined in
equation~(1)] depends only on the mass of the satellite galaxy, the
enclosed mass of the parent, and the distance between them.  To
investigate the likelihood that M32 is on such an orbit, its tidal
radius is calculated based on the following: M32 is assumed to have a
circular orbit; the projected distance between M32 and M31 is adopted
as their separation; $M_{\rm M32}=2.1\times 10^9~M_{\odot}$ is
adopted; and M31's enclosed mass is calculated by modeling it as an
isothermal sphere, $M_{\rm M31}=v_{\rm circ}^2 R_{\rm proj}/G$ with
$v_{\rm circ}=240$~km~s$^{-1}$.  The resulting $r_{\rm tide}^{\rm
M32}=310^{\sec}~$(1.2~kpc) is only weakly dependent on $M_{\rm M32}$
and $M_{\rm M31}$, so the main uncertainty is in the assumption that
$R_{\rm proj}$ is the true separation.  A measured $r_{\rm break}^{\rm
M32}=140^{\sec}~(0.54$~kpc), results in $r_{\rm break}^{\rm
M32}/r_{\rm tide}^{\rm M32}\sim0.5$, which is a conservative upper
limit since $R_{\rm proj}$ is a lower limit to the true separation.
The top panels of Figure~\ref{fig13} show the orbital eccentricity and
phase dependence of this ratio.  The ratio $r_{\rm break}/r_{\rm
tidal}$ typically has values of unity or greater for near-circular
orbits.  Only in highly eccentric orbits with $e\gtrsim0.5$ does it
drop as low as $r_{\rm break}/r_{\rm tide}\sim0.5$, suggesting that
M32 is on this latter type of orbit.

The final clue to M32's orbital eccentricity is the coincidence of
$r_{\rm break}$ to $r_{\rm distort}$, the radius associated with the
onset of isophotal elongation.  For the intrinsically-spherical,
simulated satellites, $r_{\rm distort}$ is defined as the radius at
which $\epsilon>0.02$; however, for the observations, due to the
non-spherical nature of real galaxies, $r_{\rm distort}$ is modified
to a more general definition of the radius at which the ellipticity
departs sharply from the inner radius value.  For M32 this is seen to
occur at $r_{\rm distort}^{\rm M32}=150^{\sec}~(0.57$~kpc), resulting
in $r_{\rm break}/r_{\rm distort}\sim1.0$.  In Figure~\ref{fig12}, the
locations of $r_{\rm break}$ and $r_{\rm distort}$ are shown as dotted
vertical lines in the $\Delta\mu$ and $\epsilon$ plots, respectively.
In the lower panels of Figure~\ref{fig13}, the orbital eccentricity
and phase dependencies of the ratio $r_{\rm break}/r_{\rm distort}$
indicate that $r_{\rm break}/r_{\rm distort}\ge2.0$ for near-circular
orbits and approaches unity only for the most eccentric orbits, again
supporting the theory that M32 is on such an orbit.  Unlike $r_{\rm
tidal}$, both $r_{\rm distort}$ and $r_{\rm break}$ are directly
observable, making this deduction less model dependent and more robust
than the previous one about $r_{\rm break}/r_{\rm tide}$.

%
%

In addition to constraining M32's orbital eccentricity, the
three arguments above indicate that M32 is currently in an orbital
phase away from pericenter.  In particular, the fact that the lower
limit for $r_{\rm tide}$ is a factor of two greater than both $r_{\rm
break}$ and $r_{\rm distort}$ provides robust evidence that M32 cannot
be at pericenter.  Severe tidally induced distortions are not expected
to be seen interior to the tidal radius of a satellite at pericenter,
$r_{\rm tide,\,peri}$.  Following this line of reasoning, $r_{\rm
tidal,peri}$ can be estimated using $r_{\rm break}$ and $r_{\rm
distort}$ alone.  As is shown in Figure~15 of Paper~I, $r_{\rm
tidal,peri}\approx0.5r_{\rm break}$ for $r_{\rm break}\sim r_{\rm
distort}$.  This corresponds to $r_{\rm tide,\,peri}\approx0.3$~kpc;
and translates to a M32--M31 pericenter separation of $R_{\rm
peri}\sim0.7$~kpc, via the King formula.  Even the most conservative
estimate of $r_{\rm tide,\,peri}\approx r_{\rm break}$ implies an upper
limit $R_{\rm peri}\lessapprox1.7$~kpc that is much less than $R_{\rm
proj}=5.5$~kpc.  Adopting $R_{\rm peri}=0.7$~kpc, and an orbital
eccentricity $e=0.88$ (based on Model 4 of the simulations), M32's
apocenter is estimated to be $R_{\rm apo}\approx10.5$~kpc.  If M32 is
currently near apocenter, as the $\phi^{\prime}$ triple twist
suggests, M32 must be at least $8-9$~kpc in the foreground or
background of M31's core.  This is well within the current
$\pm100$~kpc uncertainty in the relative distances to M32 and M31.  


\subsection{Direction of Motion}

For circular orbits, $\phi^{\prime}$($r_{\rm break}$) and
$d\phi^{\prime}/dr$($r_{\rm break}$) are related to the direction of
the orbit and can therefore be used to constrain the satellite's
projected motion.  Unfortunately, these relationships have a phase
dependence for eccentric orbits.  As a result, the projected motions
of M32 and NCG~205 are indeterminable from their isophote orientations
alone.  In the case of M32, the orbital direction can be recovered
since its phase has been independently determined.

If M32 is indeed on a highly eccentric orbit approaching apocenter, as
suggested in \S6.3, then the simulations indicate that $\phi_{\rm
M32}^{\prime}$($r_{\rm break}$) should be negative (Fig.~11 of
Paper~I).  The orientation of $\phi^{\prime}$ is defined with respect
to the direction of motion, implying that M32's $r_{\rm break}$
isophote, on the inner side of its orbit, should be pointed away from
its direction of motion.

\begin{figure}
\begin{center}
\epsscale{1.0}
\plotone{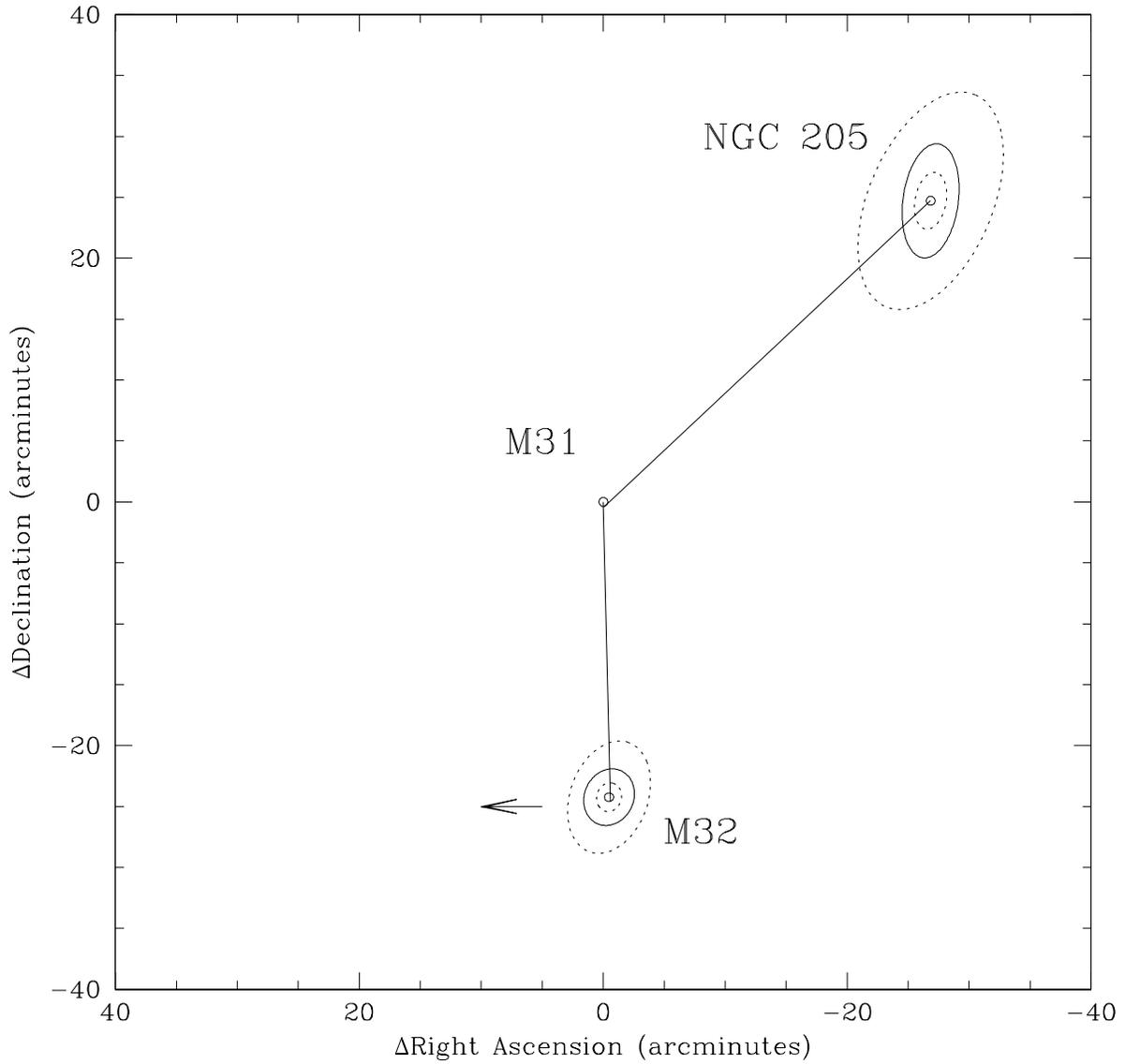}
\caption{Orientation of ellipses fit to M32 and NGC~205 relative to 
M31.  The solid ellipse is at $r_{\rm break}$ and the dotted ellipses
are at 0.5 and 2 $r_{\rm break}$.  North is up and east is left as in
Figures~\ref{fig2} and \ref{fig10}.  The arrow indicates the probable
direction of M32's projected orbit (see \S6.3).
\label{fig14}}
\end{center}
\end{figure}

One caveat of the above argument is that the non-spherical nature of
real galaxies implies that there is a non-zero intrinsic value for the
position angle, $\phi_{\rm inner}^\prime$.  Instead of simply looking
at the sign of $\phi^\prime(r_{\rm break})$, one must instead consider
the sign of the {\it change\/} in position angle relative to the
interior intrinsic value, $\Delta\phi^\prime(r_{\rm
break})\equiv\phi^\prime(r_{\rm break})-\phi_{\rm inner}^\prime$,
where $\phi_{\rm inner}^\prime=-20.0^\circ$ for M32's inner isophotes.
For M32, $\Delta\phi^\prime(r_{\rm break}$) has an absolute value of
$5.2^\circ$.  As discussed above, prior knowledge of this satellite's
orbital phase and eccentricity indicates that
$\Delta\phi^\prime(r_{\rm break}$) must be negative, and this implies
that M32's projected orbit is clockwise about M31 as indicated in
Figure~\ref{fig14}.

\subsection{Comparison of M32 and NGC~205 Intrinsic Profiles}
The generic characteristics shared by all of the simulated satellites
are a depletion zone at small radii and an excess region at large
radii (Fig.~6 of Paper~I). This is an expected consequence of tidal
stripping and flux conservation that is independent of orbital
parameters or the satellite's initial profile.  Both regions have
$\mu$ profile breaks associated with their onset: a gradual {\it
downward} (negative) break in the case of the depletion zone, and an
abrupt {\it upward} (positive) break in the case of the outer excess
region.

Though not easily discernible from the $\mu$ profiles, both breaks are
generally evident in the $\Delta\mu$ residual plots.  To mimic the
analysis of real galaxies, for which intrinsic profiles are unknown,
$r_{\rm break}$ is measured using only the $\mu$ profile.  The
detection criteria for $r_{\rm break}$ is not biased towards either
positive or negative departures; however, it does depend on the
sharpness and magnitude of the profile slope change.  As a result,
$r_{\rm break}$ tends to be preferentially associated with the sharp,
outer break.  Only in the simulated satellite of Model 5, which has a
shallow initial density profile, is the inner ``depletion zone'' break
detected as $r_{\rm break}$.  The satellites in Models 4 \& 5 have
identical orbital parameters and differ only in their initial density
profile, revealing a connection between the intrinsic profile and its
measured parameters after interaction.

In the observed $\mu$ profiles, $r_{\rm break}$ is positive for M32
and negative for NGC~205.  The difference in the intrinsic profiles of
the two satellites --- NGC~205's is much shallower than M32's ---
hints at a profile dependent detection bias, as suggested by the
simulations.  For M32, it is evident from the $\Delta\mu$ profiles
(Fig.~\ref{fig12}) that $r_{\rm break}$ corresponds to an outer
``excess region'' break.  For both M32 and the Model~4 satellite,
though not initially identified due to its gradual nature, the inner
``depletion zone'' break is clearly visible.  By contrast, because of
NGC~205's shallow intrinsic profile, the stripping of material results
in an inner ``depletion zone'' break that is sharp enough to be
measured (Fig.~\ref{fig11}).  Unfortunately, there is no evidence for
the accompanying outer ``excess region'' break that would reinforce
its identification.  This may simply be beyond the
current sensitivity limit of the observations.

\subsection{Constraints on Mass Loss}
It is shown in Paper~I that constraints can be placed on satellite mass-loss rates
from surface photometry alone.  Measurements of
the extra-tidal population of M32 and NGC~205 are used to make order-of-magnitude estimate 
for their instantaneous, fractional mass-loss rate per orbital period,

\begin{equation}
	{df \over dt} = { \pi^2 r_{\rm break}^2 \Sigma_{\rm break}
\over m_{\rm break}},
\end{equation}

\noindent as discussed in Paper~I.  The ratio of $\Sigma_{\rm
break}$/$ m_{\rm break}$, the surface density at $r_{\rm break}$ to
the mass enclosed within this point, can be calculated from the $\mu$
profile, assuming a constant mass-to-light ratio.  The derived rates
for M32 and NGC~205, $df/dt|_{\rm M32}=0.38$ and $df/dt|_{\rm
NGC~205}=2.95$, are shown in the last column of Table \ref{tab2}.

The high apparent destruction rates for both satellites should be
qualified by two factors.  The first is that although
the simulations show that surface brightness derived rates are
accurate to within order unity for near-circular orbits, this
relationship degenerates with eccentricity.  For eccentric orbits, 
$df/dt$ is phase dependent, as illustrated in
Figure~16 of Paper~I.  Mass loss estimates are reliable near
pericenter, where the bulk of mass loss occurs; however, away from this phase,
they are systematically high by up to half an order of
magnitude.  The reason for this overestimate is that away from
pericenter, only a fraction of the extra-break material that is
generally heated, yet bound, will be lost on the current orbit.  The
second factor is that $df/dt$ is an instantaneous, phase-dependent rate that
provides a direct measure of the total mass loss per orbit only when $df/dt$ 
is constant, as in the circular case.  For eccentric orbits, it must be 
integrated over the entire orbit to calculate the total orbital mass loss.
Given these caveats, the presented $df/dt$ should be
considered only as upper limits for the instantaneous fractional
mass-loss rate.  As such, they should not be used to extrapolate a 
destruction rate.

\subsection{Future Directions}

The simulations presented in this paper and Paper~I provide
useful pointers about the nature of tidal interaction in M32 and NGC~205,
despite the fact that they are not tuned to mimic these satellites.  Future
simulations will explore combinations of satellite orbital eccentricities and
phases that are constrained by the actual distances and radial velocities
measured for M32, NGC~205 and M31.  Furthermore, the simulated satellites
almost certainly depart from real galaxies in the assumption that mass
follows light.  In the future, two-component (stars and dark matter)
model satellites will be incorporated into the simulations.

\section{Implications for M32's Surface Brightness and Luminosity Evolution}
In a plot of $L$ versus $\mu$, cEs lie on the extension of the giant E
galaxy track, typically $\sim$$2--3$ magnitudes fainter in luminosity
and $\sim$$1--2$ magnitudes brighter in surface brightness \citep{zb98}.
At the faint, low surface brightness extreme is M32.  Due to its
proximity to M31, most formation theories suggest that M32 is the
remnant of a galaxy that has been stripped through tidal interaction.
Numerous galaxy types have been proposed as possible precursors; however,
given its location in $\mu$-$L$ space, the most intuitive of these is
a normal E galaxy.  This theory can be investigated directly using our
surface brightness observations.

The simulations presented in \S6 indicate that, despite the loss of 
a substantial amount of mass in their outer regions, the
interior portions of the dwarf satellites' $\mu$ profiles
remain largely unaffected [(Fig.~\ref{fig12} (upper two panels on the
right)].  While the
simulations are admittedly simplistic, in the case of 
M32, this assumption is probably a reasonable one.  One particular concern is that the
present simulations involve single-component satellites in which mass
follows light.  By contrast, real satellites are expected to have extended
dark halos.  Fortunately, such a halo would tend
to further buffer the interior of the satellites from tidal stripping, thereby 
reinforcing this finding.

Guided by the notion that the interior brightness profile of M32 is
pristine and the {\it assumption\/} that the original profile (prior
to tidal stripping) obeyed a de~Vaucouleurs law, one can
quantitatively address the question of whether the unusual location of
this galaxy in a $\mu$-$L$ plot could be the result of tidal stripping
of a normal E galaxy.  Of the three $r^{1/4}$ law fits presented in
Figure~\ref{fig5} of \S4.3, the one labeled ``extreme-inner'' (fit to
inner $r=10''$--$30''$) is most likely to represent the intrinsic
profile of M32.  The resulting estimates of M32's intrinsic effective
surface brightness are $\mu_I^{\rm
eff}=18.41$~mag~arcsec$^{-2}$ and $\mu_B^{\rm
eff}=20.15$~mag~arcsec$^{-2}$.  Adopting the ``standard''
($r_{\rm eff}=29^{\sec}$) fit as representative of M32's current
de~Vaucouleurs law profile leads to current M32 values of $\mu_I^{\rm
eff}=17.53$ and $\mu_B^{\rm eff}=19.43$, in good agreement with
historical results.  These values imply an evolution of $\Delta
\mu_I^{\rm eff}=0.88$ and $\Delta \mu_B^{\rm eff}=0.72$.

\begin{figure}
\begin{center}
\epsscale{1.0}
\plotone{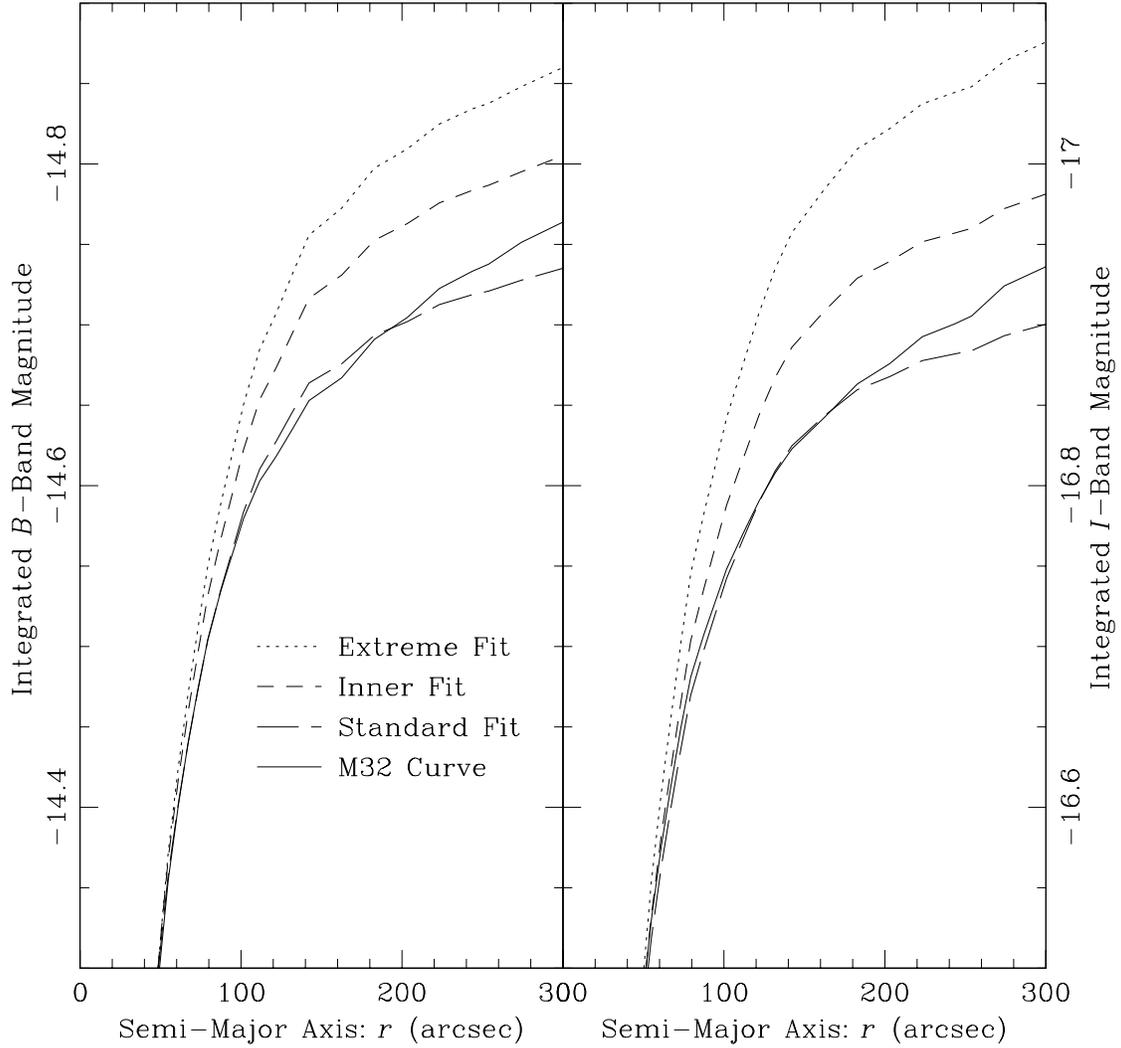}
\caption{Absolute magnitude, based on integrated light within an
isophote, as a function of isophotal radius (``curve of growth'') for
M32 in $B$ ({\it left\/}) and $I$ ({\it right\/}) bands.  The growth
curve for the measured M32 profile ({\it solid line\/}) is shown in
contrast to curves based on integration of the standard, inner and
extreme-inner ({\it long dashed, short dashed, dotted lines\/})
$r^{1/4}$ law fits listed in Table~1.  Adopting either the inner or
extreme-inner fit as M32's intrinsic profile suggests that its
luminosity (within the $r=300^{\sec}$ isophote) has evolved by $\Delta
B\sim0.05\>$--$\>0.10$ and $\Delta I\sim0.05\>$--$\>0.15$.
\label{fig15}}
\end{center}
\end{figure}

The luminosity evolution is estimated by comparing its current
luminosity to that of the intrinsic $r^{1/4}$ law profile fit assumed
for M32.  The three de~Vaucouleurs law fits are integrated to estimate
the intrinsic luminosity of M32 as a function of enclosed radius.
These ``curves of growth'' are shown in Figure~\ref{fig15}
(long-dashed, short-dashed, and dotted lines for the ``standard'',
``inner'', and extreme-inner'' fits, respectively), along with the
curve of growth based on the {\it actual\/} brightness profile of M32
(solid line).

These curves overlap with one another interior to the inner break
$r=50^{\sec}$ and diverge beyond this radius, consistent with the
expectation that the majority of the luminosity evolution occurs in
the depletion zone.  The ``standard fit'', which most closely follows
M32's observed integrated luminosity curve through the depletion zone,
underestimates the total magnitude at large radii, due to the tidal
excess feature discussed in \S4.2.  The curves of the two inner-radius
fits on the other hand are less biased by the depletion zone and
therefore provide a more conservative estimate for M32's intrinsic
luminosity.  Adopting the ``extreme-inner'' fit as M32's intrinsic
profile results in a modest luminosity evolution of $\Delta B_{\rm
M32}\sim$$0.1$ and $\Delta I_{\rm M32}\sim$$0.15$, based on aperture
photometry out to $r=300^{\sec}$.  The adoption of total versus
isophotal magnitudes would only impact the luminosity evolution by
an additional $\sim$$10\%$.

A comparison of M32's presently observed properties to estimates of its
intrinsic properties, indicates a relatively small amount of
evolution due to tidal effects.  Although it is in the right direction
--- away from the family of E galaxies in the $\mu$-$L$ projection
--- the magnitude of this shift falls far short of explaining M32's
position in terms of a tidally stripped/truncated normal E galaxy.
Put another way, intermediate E galaxies have typical effective radii
of $1.2<r_{\rm eff}<8.0$~kpc \citep{bbf93}, whereas estimates of M32's
intrinsic effective radius are in the range $37^{\sec}$--$47^{\sec}$
(0.14--0.18~kpc).  This implies that M32 was intrinsically `compact'
even before any tidal stripping by M31, supporting Burkert's (1994a)
theory that cEs are {\it formed\/} in a compact state --- as opposed to
being evolved into one.  The bulges of spiral or S0 galaxies,
typically intermediate in compactness between Es and cEs,
cannot be ruled out based on  the current analysis.

It is interesting to note that the integrated absolute magnitude of the
\citet{ibata+01} stream, estimated at $M_V({\rm stream})\approx-14$, is
approximately 10\% that of M32: $M_V({\rm M32})\lesssim-16$.  The M32 value
is derived from its curve of growth in the $B$ band which yields $M_B({\rm
M32})\lesssim-15$ (Fig.~\ref{fig15}), and an interpolated $B-V$ color of
about unity (see \S4.4.2).  Moreover, the estimated amount of luminosity
evolution in M32 due to tidal stripping is about 0.1~mag (see above).  Thus
accumulated tidal debris from M32 can adequately account for the overall
brightness of the stream.

\section{Summary}

This paper presents surface photometry of M31's two nearest satellites,
M32 and NGC~205, and a comparison to N-body simulations.  Details of the
simulations are in the companion paper Johnston, Choi \& Guhathakurta
(Paper~I).  The primary objectives of this work are to investigate the
impact of tidal interactions on the morphology and evolution of dwarf
satellite galaxies and to place constraints on the satellite orbital
parameters.  The main points are outlined below:

\begin{itemize}

\item{Large-format $B$- and $I$-band CCD mosaic images of the M31 sub-group
form the basis of this study.  Global ellipse fits are used to model and
subtract M31's contaminating disk light, enabling measurement of the faint
outer isophotes of M32 and NGC~205 where tidal signatures are most prominent.}

\item{The surface brightness profile of M32 has traditionally been fit by a
de~Vaucouleurs $r^{1/4}$ law, but there is a clear excess of light in the
outer parts ($r\gtrsim140^{\sec}$) relative to the ``standard'' fit.  The
excess is coincident with elongation and twisting of the isophotes.  There is
also a downward break in the $\mu$ profile at $r\sim50^{\sec}$ in the inner
region of M32; this too is accompanied by isophote twists.  The intrinsic
$\mu$ profile of NGC~205 is more complex than M32's, intermediate between a
simple exponential and $r^{1/4}$ laws, and is in good agreement with previous
measurements.}

\item{The robustness of the M32 results is demonstrated through a series of
tests.  The measured isophotal parameters---surface brightness, ellipticity,
and orientation---are robust out to at least $r\sim250^{\sec}$ and share the
following characteristics: (1)~insensitive to details of M31 disk modeling
and sky subtraction errors; (2)~symmetric about M32 despite the stark
difference in the quality of the inner versus outer M31 disk list
subtraction; and (3)~$B-I$ color index that is consistent with the inner
parts of M32.}

\item{The M32 and NGC~205 measurements are compared to numerical simulations
of single-component, spherical, non-rotating, satellites, orbiting in a
fixed, three-component parent galaxy potential.  The simulations provide
insight into the nature of tidal interaction even though they are not
tailored precisely to M32 and NGC~205.}

\item{The surface brightness profiles of tidally disrupted simulated
satellites contain certain generic features reminiscent of those seen in the
M31 satellites.  These features include an excess region at large radii, a
depletion zone at intermediate radii, and a central region that is largely
unaffected by tidal interaction.  Isophote elongation and twists are also
common, though the details of the $\epsilon$ and $\phi^{\prime}$ radial
profiles are strongly dependent on orbital phase.}

\item{A comparison between the observations and numerical simulations
indicates that M32 and NGC~205 are likely both on highly eccentric orbits,
away from pericenter, and that they are being viewed from outside their
orbital plane.  The sense of M32's projected orbit around M31 appears to be
clockwise.  M32 has a simpler (intrinsic) brightness distribution in its
inner parts than NGC~205 and is a better match to our current suite of
simulations; its orbital parameters are therefore better constrained.}

\item{Empirical estimates are made of the effect of tidal stripping on M32's
luminosity and effective surface brightness, based on an extrapolation of its
inner surface brightness profile.  The estimated amount of change in $L$ and
$\mu_{\rm eff}$ is far too small to be consistent with the theory that M32
evolved from a normal elliptical, and suggests instead that M32's precursor
was intrinsically more compact than a typical E galaxy.  This supports
Burkert's (1994a) formation scenario for compact ellipticals such as M32
through a starburst and subsequent violent collapse within the potential well
of a massive galaxy, though spiral or S0 bulges cannot be ruled out as
possible precursors.}

\item{While the current numerical simulations provide qualitative insight
into the nature of tidal interaction in the M31 sub-group, future simulations
will be tailored specifically to match the observed radial velocities,
line-of-sight distances, and dynamical masses of M31, M32, and NGC~205.
Other planned improvements include two-component satellites (stars and dark
matter).  Keck spectroscopy of individual red giant stars in the tidal region
of M32 is being used to measure velocity and velocity dispersion profiles,
which should better constrain the details of its interaction with M31.}

\end{itemize}

\section{Acknowledgements}
We thank the referee for her/his many detailed and insightful
comments.  We would like to acknowledge the help and insight of Somak
Raychaudhury, our collaborator on the M31 CCD mosaic observations, for
making this project possible in the first place.  We are grateful to
Mike Bolte and Sandy Faber for useful discussion and to Patrik Jonsson
and Anouk Shambrook for careful readings of the manuscript.  PIC
thanks the ARCS foundation and the NSF for support as an ARCS
Foundation scholar and an NSF graduate student research fellow.  KVJ
acknowledges support in part as a member of the Institute for Advanced
Study, and from NASA LTSA grant NAG5-9064.

\newpage


\newpage
\clearpage

\begin{deluxetable}{ccccccc} 
\tablecolumns{7} 
\tablewidth{0pc} 
\tablecaption{M32 de~Vacouleurs Profile Fit Parameters} 
\tablehead{\colhead{Band} & \colhead{${r_{\rm inner}}$} &
\colhead{${r_{\rm outer}}$} & \colhead{$r_{\rm eff}$} &
\colhead{$\mu_{\rm eff}$} & \colhead{$\Delta\mu_{\rm eff}$} &
\colhead{Comments}\\ 
\colhead{} & \colhead{($^{\sec}$)} &
\colhead{($^{\sec}$)} & \colhead{($^{\sec}$)} & \colhead{(mag)} &
\colhead{(mag)}} 
\startdata 
$R$ & 15 & 100 & 32.0 & 18.79 & --- & Kent (1987) Data\\ 
$R$ & 10 & 140 & 32.5 & 18.64 & --- & Kent (1987) Data\\
$I$ & 10 & 140 & 28.5 & 17.53 & --- & Standard Fit\\ 
$B$ & 10 & 140 & 28.5 & 19.43 & --- & Standard Fit\\ 
$I$ & 10 & 65 & 36.8 & 18.00 & 0.47 & Inner Fit\\ 
$B$ & 10 & 65 & 36.4 & 19.90 & 0.47 & Inner Fit\\
$I$ & 10 & 30 & 46.8 & 18.41 & 0.88 & Extreme-Inner Fit\\ 
$B$ & 10 & 30 & 42.0 & 20.15 & 0.72 & Extreme-Inner Fit\\ 
\enddata
\label{tab1}
\end{deluxetable} 

\begin{deluxetable}{ccccccccc} 
\tablecolumns{8} 
\tablewidth{0pc} 
\tablecaption{Observed Profile Parameters and Derived Quantities for M32 \& NGC~205} 
\tablehead{ \colhead{Name} & \colhead{$R_{\rm proj}$\tablenotemark{a}} &
\colhead{$r_{\rm tide}$\tablenotemark{a}} &
\colhead{$r_{\rm break}$\tablenotemark{a}} & 
\colhead{$r_{\rm distort}$\tablenotemark{a}} &
\colhead{$\epsilon$(${\rm break}$)}& 
\colhead{$\phi^{\prime}$(${\rm break}$)} & \colhead{$r d\phi^{\prime}/dr$($r_{\rm break}$)}&
\colhead{$df/dt$} \\
\colhead{} & \colhead{(kpc)} & \colhead{(kpc)} & \colhead{(kpc)}&
\colhead{(kpc)} & \colhead{} & \colhead{} & \colhead{} &
\colhead{}}
\startdata 
M32 	 & 5.5 & 1.2 & 0.54 & 0.57 & 0.14 &$-25.24$ & 12.60 & 0.38\\
NGC~205  & 8.3 & 1.0 & 1.07 & ---  & 0.52 &$-36.32$ & 66.26 & 2.95\\
\enddata 
\tablenotetext{a}{Based on an assumed distance to M31, M32, and NGC~205 of 780~kpc}

\label{tab2}
\end{deluxetable} 

\end{document}